\documentclass[aps,prd]{revtex4}

\usepackage{amsmath}
\usepackage{amssymb}
\usepackage{bm}% bold math
\usepackage{graphicx}
\usepackage{multirow}
\usepackage{color,soul}

\usepackage{textcomp}

\allowdisplaybreaks[1]

\begin{document}
%\preprint{  }

\title{Dispersive determination of fourth generation quark masses}

\author{Hsiang-nan Li}
\affiliation{Institute of Physics, Academia Sinica,
Taipei, Taiwan 115, Republic of China}

\date{\today}

\begin{abstract}

We determine the masses of the sequential fourth generation quarks $b'$ and $t'$ in the
extension of the Standard Model by solving the dispersion relations associated with the 
mixing between the neutral states $Q\bar q$ and $\bar Qq$, $Q$ ($q$) being a heavy 
(light) quark. The box diagrams responsible for the mixing, which provide the 
perturbative inputs to the dispersion relations, involve multiple intermediate channels, 
i.e., the $ut$ and $ct$ channels, $u$ ($c$, $t$) being an up (charm, top) quark, in 
the $b'$ case, and the $db'$, $sb'$ and $bb'$ ones, $d$ ($s$, $b$) being a down 
(strange, bottom) quark, in the $t'$ case. The common solutions for the above channels 
lead to the masses $m_{b'}=(2.7\pm 0.1)$ TeV and $m_{t'}\approx 200$ TeV unambiguously. We 
show that these superheavy quarks, forming bound states in a Yukawa potential, barely 
contribute to Higgs boson production via gluon fusion and decay to photon pairs, and 
bypass current experimental constraints. The mass of the $\bar b'b'$ ground state
is estimated to be about 3.2 TeV. It is thus worthwhile to continue the search for
$b'$ quarks or $\bar b'b'$ resonances at the (high-luminosity) large hadron collider.

\end{abstract}

\maketitle

%Unitarity bound, M.S. Chanowitz, M.A. Furman and I. Hinchliffe, Phys.
%Lett. B 78, 285 (1978); Nucl. Phys. B 153, 402 (1979).

%--------+---------+---------+---------+---------+---------+---------+---------+
\section{INTRODUCTION}

Our recent dispersive analyses of some representative physical observables (heavy meson 
decay widths, neutral meson mixing, etc.) have accumulated substantial indications that the 
scalar sector of the Standard Model (SM) is not completely free, but arranged properly to 
achieve internal dynamical consistency \cite{Li:2023dqi,Li:2023yay,Li:2023ncg}. Fermion 
masses can be derived by solving the dispersive relations for decay widths of a  
heavy quark $Q$ as an inverse problem \cite{Li:2020xrz,Li:2020fiz,Li:2020ejs,Xiong:2022uwj}: 
starting with massless final-state up and down quarks, we demonstrated that the solution for 
the $Q\to du\bar d$ ($Q\to c\bar ud$) mode with the leading-order heavy-quark-expansion input 
yields the charm-quark (bottom-quark) mass $m_c=1.35$ ($m_b= 4.0$) GeV \cite{Li:2023dqi}. 
Requiring that the dispersion relation for the $Q\to su\bar d$ ($Q\to d\mu^+\nu_\mu$, 
$Q\to u\tau^-\bar\nu_\tau$) decay generates the identical heavy quark mass, we deduced the 
strange-quark (muon, $\tau$ lepton) mass $m_s= 0.12$ GeV ($m_\mu=0.11$ GeV, $m_\tau= 2.0$ GeV). 
The similar studies of fermion mixing \cite{Li:2023ncg} established the connections between 
the Cabibbo-Kobayashi-Maskawa (CKM) matrix elements and quark masses, and between the 
Pontecorvo–Maki–Nakagawa–Sakata matrix elements and neutrino masses. These connections 
explained the known numerical relation $V_{us}\approx \sqrt{m_s/m_b}$ \cite{Belfatto:2023qca}, 
$V_{us}$ being a CKM matrix element, and the maximal mixing angle $\theta_{23}\approx 45^\circ$ 
in the lepton sector, and discriminated the normal hierarchy for neutrino masses from the 
inverted hierarchy or quasi-degenerate spectrum.

The dispersion relation for the correlation function of two $b$-quark scalar (vector) 
currents, with the perturbative input from the $b$ quark loop, returns the Higgs ($Z$) boson 
mass 114 (90.8) GeV \cite{Li:2023yay} in accordance with the measured values. It implies that 
the parameters $\mu^2$ and $\lambda$ in the Higgs potential are also constrained by internal 
dynamical consistency. Particle masses and mixing angles in the SM originate 
from the independent elements of the Yukawa matrices \cite{Santamaria:1993ah}, as the 
electroweak symmetry is broken. Inspired by the above observations, we attempt to make a bold 
conjecture that the SM contains only three fundamental parameters actually, i.e., the three 
gauge couplings, and the other parameters, governing the interplay among various generations 
of fermions, are fixed by SM dynamics itself. The analyticity, which is inherent in quantum 
fields theories, imposes additional constraints. Its impact is not revealed in 
naive parameter counting at the Lagrangian level based on symmetries, but through dispersive 
analyses of dynamical processes. Dispersion relations, which physical observables like 
heavy-to-light decay widths must respect, link different types of interactions at arbitrary 
scales. The resultant constraints are so strong that the parameters in the 
scalar sector must take specific values, instead of being discretionary.

To maintain the simplicity and beauty conjectured above, a natural extension of the SM is to 
introduce the sequential fourth generation of fermions, since the associated parameters in 
the scalar sector are not free. That is, their masses and mixing with lighter generations 
can be predicted unambiguously in a similar manner \cite{Li:2023yay}. We first 
determine the top quark mass $m_t$ by solving the dispersion relations for 
the mixing between the neutral states $Q\bar u$ and $\bar Qu$. The perturbative inputs to the 
dispersion relations come from the imaginary contributions of the box diagrams for 
the mixing with the intermediate $db$, $sb$ and $bb$ channels. Given the corresponding 
thresholds $m_d+m_b$, $m_s+m_b$, and $2m_b$ for the typical quark masses $m_d=0$, $m_s=0.1$ 
GeV, and $m_b=(4.15\pm 0.01)$ GeV, we extract $m_t=(173\pm 3)$ GeV from the common solution 
to the three channels. The existence of such a common solution is highly nontrivial, 
making convincing our formalism and predictions obtained from it. We then go ahead 
to calculate the masses of the sequential fourth generation quarks $b'$ and $t'$ in the same 
framework, considering the multiple intermediate channels $ut$ and $ct$ in the $b'$ case, and 
$db'$, $sb'$ and $bb'$ in the $t'$ case. It will be observed that the common solutions for the 
various channels also exist, and demand the masses $m_{b'}=(2.7\pm 0.1)$ TeV and 
$m_{t'}\approx 200$ TeV.

Many merits of the sequential fourth generation model have been explored: condensates of the 
fourth generation quarks and leptons could be the responsible mechanism of the dynamical 
electroweak symmetry breaking \cite{HBH,Mimura:2012vw}; electroweak baryogenesis through the 
first-order phase transition could be realized in this model \cite{HOS}; it could provide a 
viable source of $CP$ violation for the baryon asymmetry of the Universe based on the dimensional 
analysis of the Jarlskog invariants \cite{Hou:2008xd}. However, it is widely conceded that this 
model has been ruled out mainly by the data of Higgs boson production via gluon fusion $gg\to H$ 
and decay into photon pairs $H\to\gamma\gamma$ \cite{Chen:2012wz}. Measurements of the oblique 
parameters, which depend on the additional mixing angles associated with the 
fourth generation quarks and the unclear contribution from the fourth generation leptons 
\cite{He:2001tp}, give relatively weaker constraints. We point out that the superheavy fourth 
generation quarks $b'$ and $t'$ with the aforementioned masses form bound states in a Yukawa 
potential \cite{Hung:2009hy,Enkhbat:2011vp}. Once 
they form bound states, physical degrees of freedom change, and new 
resonances emerge, so one has to reformulate the interaction between the fourth generation 
quarks and Higgs bosons with these new resonances \cite{Hung:2010xh}. We will show 
that the $\bar b'b'$ scalars contribute to the $gg\to H$ cross section only at 
$10^{-3}$ level, relative to that from the top-quark loop in the SM. It is thus 
likely for the sequential fourth generation model to bypass the current experimental 
constraints, even without the expansion of the scalar sector \cite{Das:2017mnu}. For an analogous 
reason, the model could also bypass the constraint from Higgs boson decay to photon pairs.

%, such as the three-Higgs-doublet potential proposed in \cite{Hung:2010xh}

The rest of the paper is organized as follows. We compute the 
top quark mass from the dispersion relations for the $Q\bar u$ and $\bar Qu$ mixing through 
the $db$, $sb$ and $bb$ channels in Sec.~II. The framework is extended to the prediction 
for the $b'$ ($t'$) quark mass in Sec.~III by investigating the multiple 
intermediate $ut$ and $ct$ ($db'$, $sb'$ and $bb'$) channels. The properties of the $\bar b'b'$ 
scalar bound states $S$ in a Yukawa potential, including the binding energies and the widths, 
are derived in Sec.~IV, based on which we estimate the $gg\to S\to H$ cross sections using 
the $ggS$ and $SH$ effective couplings and Breit-Wigner propagators for $S$. In particular,
the mass of the $\bar b'b'$ ground state, being either a pseudoscalar or a vector, is evaluated 
in a relativistic approach and found to be about 3.2 TeV. Some processes, which are promising 
for searching for $b'$ quarks and their resonances at the (high-luminosity) large hadron 
collider, are proposed. Section V contains the summary.

\section{FORMALISM AND TOP QUARK MASS}

\begin{figure}
\begin{center}
\includegraphics[scale=0.35]{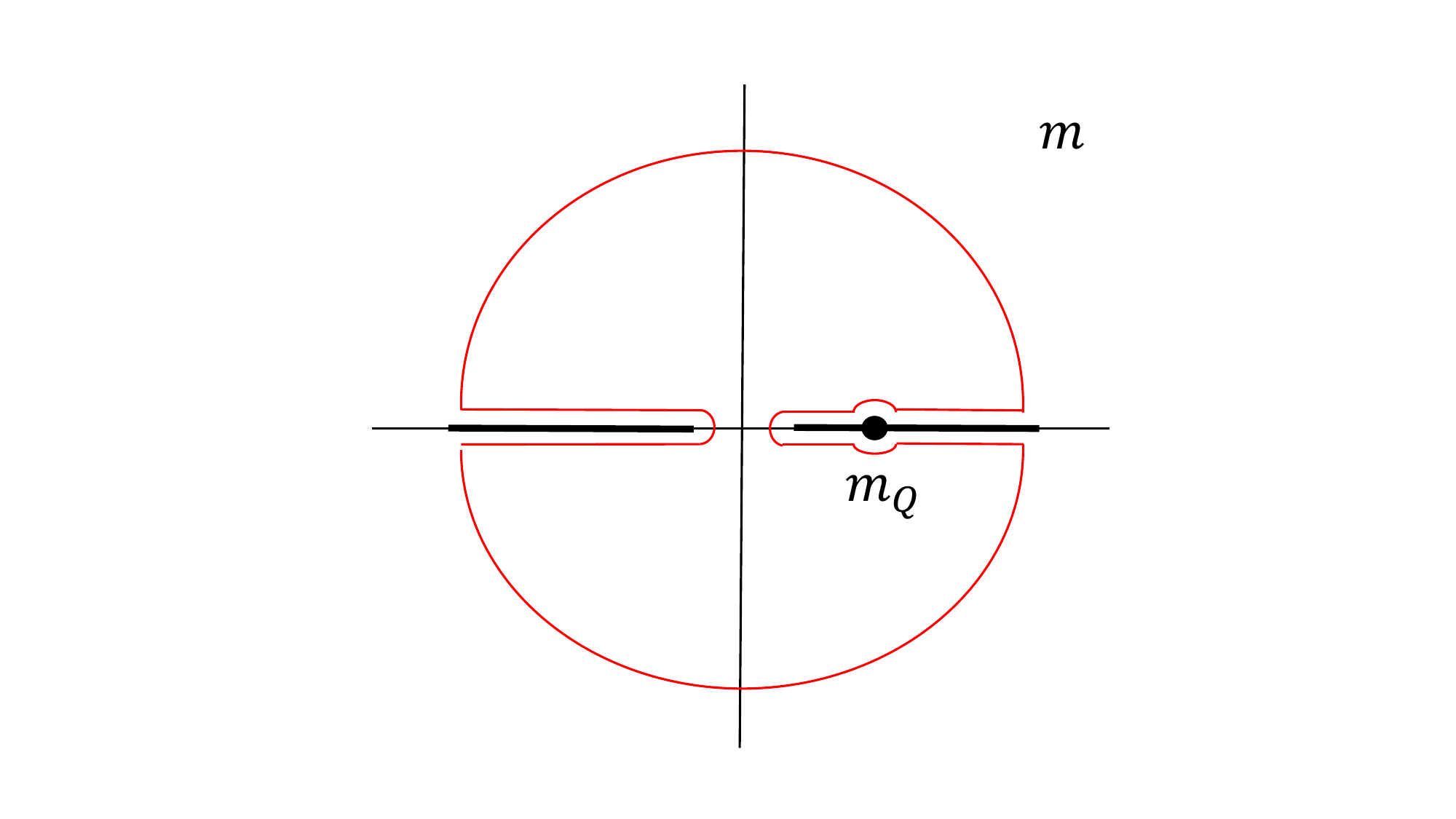}
\caption{\label{fig1}
Contour for the derivation of Eq.~(\ref{it0}), where the thick lines represent the branch cuts.}
\end{center}
\end{figure}

Consider the mixing between the neutral states $Q\bar u$ and $\bar Qu$ through the box 
diagrams with a heavy quark $Q$ of mass $m_Q$ and a massless $u$ quark 
\cite{Li:2022jxc,Li:2023yay}. The construction of a dispersion relation follows the procedure 
in \cite{Li:2023dqi} straightforwardly, which starts with the contour integration of the mixing 
amplitudes $\Pi_{ij}$, $ij=db$, $sb$ and $bb$, in the complex $m$ plane. The contour consists 
of two pieces of horizontal lines above and below the branch cut along the positive real axis, 
two pieces of horizontal lines above and below the branch cut along the negative real axis, a 
small circle around the pole $m=m_Q$ located on the positive real axis and 
a circle $C_R$ of large radius $R$ as depicted in Fig.~\ref{fig1}. As
recollected in Appendix A, we have the dispersion relations for the imaginary pieces of $\Pi_{ij}$
\begin{eqnarray}
\int_{M_{ij}^2}^{R^2}\frac{{\rm Im}\Pi_{ij}(m)}{m_Q^2-m^{2}}dm^2
=\int_{m_{ij}^2}^{R^2}\frac{{\rm Im}\Pi^{\rm box}_{ij}(m)}{m_Q^2-m^{2}}dm^2.\label{it0}
\end{eqnarray}
The quark-level thresholds $m_{ij}$ for the box-diagram contributions ${\rm Im}\Pi^{\rm box}_{ij}$ 
denote $m_i+m_j$, i.e., $m_{db}=m_d+m_b$, $m_{sb}=m_s+m_b$ and $m_{bb}=2m_b$. The physical 
quantities ${\rm Im}\Pi_{ij}(m)$ on the left-hand side of the above expression have the hadronic 
thresholds $M_{db}=m_{\pi}+m_B$, $M_{sb}=m_K+m_B$ and $M_{bb}=2m_B$ with the pion (kaon, $B$ meson) 
mass $m_\pi$ ($m_K$, $m_B$). The CKM factors associated with the $db$, $sb$, and $bb$ 
channels can vary independently in a mathematical viewpoint, so their corresponding dispersion 
relations can be analyzed separately. These dispersion relations, holding for arbitrary $m_Q$, 
impose stringent connections between high-mass and low-mass behaviors of the mixing amplitudes.

The box diagrams generate two effective four-fermion operators of the $(V-A)(V-A)$ and 
$(S-P)(S-P)$ structures. Viewing that the two structures endow separate dispersion relations,
and the latter also receives contributions from amplitudes other than the 
box diagrams, like the double penguin amplitude \cite{Petrov:1997ch}, 
we concentrate on the former. The imaginary piece of the $(V-A)(V-A)$ structure
in perturbative evaluations \cite{Cheng,BSS} is written as
\begin{eqnarray}
\Gamma^{\rm box}_{ij}(m_Q)&\propto&\frac{C^2_2(m_Q)}{m_Q^4}
\frac{\sqrt{[m_Q^2-(m_i+m_j)^2][m_Q^2-(m_i-m_j)^2]}}
{(m_W^2-m_i^2)(m_W^2-m_j^2)}\nonumber\\
& &\times\Bigg\{2\left(m_W^4+\frac{m_i^2m_j^2}{4}\right)
[m_Q^2-(m_i+m_j)^2][m_Q^2-(m_i-m_j)^2]\nonumber\\
& &\hspace{0.7cm}-3m_W^2m_Q^2(m_i^2+m_j^2)(m_Q^2-m_i^2-m_j^2)\Bigg\},
\label{bij}
\end{eqnarray} 
with the $W$ boson mass $m_W$ and the intermediate quark masses $m_i$ and $m_j$. A $d$ quark 
is also treated as a massless particle, i.e., $m_d=0$. The overall coefficient, 
irrelevant to the derivation below, is implicit. We have kept only the Wilson coefficient 
$C_2(\mu)$ \cite{Buchalla:1995vs}, which dominates over $C_1(\mu)$ at the renormalization scale 
$\mu=m_Q \geq m_b$. The second term in the curly brackets of Eq.~(\ref{bij}) is down by a tiny ratio 
$(m_i^2+m_j^2)/m_W^2$, so the behavior of Eq.~(\ref{bij}) in $m_Q$ is dictated by the first term. 
In the threshold regions with $m_Q\sim m_{ij}$, it is approximated by
\begin{eqnarray}
\Gamma^{\rm box}_{db}(m_Q)&\sim &\frac{(m_Q^2-m_b^2)^3}{m_Q^4},\nonumber\\
\Gamma^{\rm box}_{sb}(m_Q)&\sim&\frac{\sqrt{[m_Q^2-(m_b+m_s)^2][m_Q^2-(m_b-m_s)^2]}^3}
{m_Q^4},\nonumber\\
\Gamma^{\rm box}_{bb}(m_Q)&\sim &\frac{\sqrt{m_Q^2-4m_b^2}^3}{m_Q}.\label{asy}
\end{eqnarray}
Because of $m_s\ll m_b$, $m_b-m_s$ is not very distinct from $m_b+m_s$, and the dependence
on the former has been retained in the second line of Eq.~(\ref{asy}).

Motivated by the above threshold behaviors, we choose the integrands for the 
dispersion integrals in Eq.~(\ref{it0}) as \cite{Li:2023yay} 
\begin{eqnarray}
{\rm Im}\Pi_{db}(m)&=&\frac{m^4\Gamma_{db}(m)}{(m^2-m_b^2)^2},\nonumber\\
{\rm Im}\Pi_{sb}(m)&=&\frac{m^4\Gamma_{sb}(m)}{[m^2-(m_b+m_s)^2]^2\sqrt{m^2-(m_b-m_s)^2}^3},
\nonumber\\
{\rm Im}\Pi_{bb}(m)&=&\frac{m\Gamma_{bb}(m)}{m^2-4m_b^2},\label{mi}
\end{eqnarray}
where $\Gamma_{ij}(m)$ are the unknowns to be solved for shortly, and the definitions of 
${\rm Im}\Pi^{\rm box}_{ij}(m)$ by means of $\Gamma^{\rm box}_{ij}(m)$ should be self-evident. 
Note that $\Gamma^{\rm box}_{bb}(m)$ is an odd function in $m$, which accounts for the odd power 
of $m$ in the numerator of ${\rm Im}\Pi_{bb}(m)$ \cite{Li:2023yay}. The above integrands with 
powers of $m$ in the numerators suppress any residues in the low $m$ region, including those from 
the poles at $m=\pm(m_i+m_j)$ and $m=\pm(m_i-m_j)$, compared to the ones from 
$m=\pm m_Q$ at large $m_Q$. The denominators alleviate the divergent behaviors caused by 
the modified numerators at large $m$. The factor $\sqrt{m^2-(m_b-m_s)^2}$ in 
$\Pi_{sb}(m)$ introduces an additional branch cut along the real axis in the interval 
$-(m_b-m_s)<m< m_b-m_s$ in the $m$ plane. Our contour crosses the real axis between 
$m=-(m_b+m_s)$ and $m=-(m_b-m_s)$ and between $m=m_b-m_s$ and $m=m_b+m_s$, and runs along 
the real axis marked by $m<-(m_b+m_s)$ and $m>m_b+m_s$, such that this 
additional branch cut does not contribute.

Moving the integrands on the right-hand side of Eq.~(\ref{it0}) to the left-hand side,  
we arrive at
\begin{eqnarray}
& &\int_{m_{ij}^2}^\infty\frac{\Delta\rho_{ij}(m)}{m_Q^2-m^{2}}dm^2=0,\label{it}
\end{eqnarray}
with the subtracted unknown functions 
$\Delta\rho_{ij}(m)\equiv {\rm Im}\Pi_{ij}(m)-{\rm Im}\Pi^{\rm box}_{ij}(m)$.
Owing to the subtraction of the box-diagram contributions and the limits
${\rm Im}\Pi_{ij}(m)\to {\rm Im}\Pi^{\rm box}_{ij}(m)$ at large $m$, the integrals in 
Eq.~(\ref{it}) converge even after the upper bound of $m^2$ is extended to infinity. The 
unknowns $\Delta\rho_{ij}(m)$ are fixed to the initial conditions 
$-{\rm Im}\Pi^{\rm box}_{ij}(m)$ in the interval $(m_{ij},M_{ij})$ of $m$, in which the 
physical quantities ${\rm Im}\Pi_{ij}(m)$ vanish. The idea behind 
our formalism is similar to that of QCD sum rules \cite{SVZ}, but with power corrections in 
$(M_{ij}-m_{ij})/m_Q$ arising from the difference between the quark-level and hadronic thresholds, 
which are necessary for establishing a physical solution \cite{Li:2022jxc}. As seen later,
it is easier to solve for $\Delta\rho_{ij}(m_Q)$ than for 
$\Delta\Gamma_{ij}(m_Q)\equiv \Gamma_{ij}(m_Q)-\Gamma_{ij}^{\rm box}(m_Q)$, because the
initial conditions of the former are simpler. Once $\Delta\rho_{ij}(m_Q)$ are attained, we 
convert them to $\Delta\Gamma_{ij}(m_Q)$ following Eq.~(\ref{mi}).
Without the power corrections, i.e., if $m_{ij}$ are equal to $M_{ij}$, there will be only the 
trivial solutions $\Delta\Gamma_{ij}(m_Q)=0$, i.e., $\Gamma_{ij}(m_Q)=\Gamma^{\rm box}_{ij}(m_Q)$ 
and no constraint on the top quark mass.

The steps of solving Eq.~(\ref{it}) have been elucidated in \cite{Li:2023dqi} and briefly 
reviewed in Appendix A. The solution of the unknown function can be constructed with a single 
Bessel function of the first kind $J_\alpha(x)$,
\begin{eqnarray}
\Delta \rho_{ij}(m_Q)\approx y_{ij}\left(\omega \sqrt{m_Q^2-(m_i+m_j)^2}\right)^{\alpha_{ij}} 
J_{\alpha_{ij}}\left(2\omega \sqrt{m_Q^2-(m_i+m_j)^2}\right).\label{d2}
\end{eqnarray}
A solution to the dispersion relation must not be sensitive to the arbitrary scale $\omega$, 
which results from scaling the integration variable $m^2$ in Eq.~(\ref{it}) artificially
\cite{Li:2023yay}. To realize this insensitivity, we make a Taylor expansion of 
$\Delta \rho_{ij}(m_Q)$,
\begin{eqnarray}
\Delta \rho_{ij}(m_Q)=\Delta \rho_{ij}(m_Q)|_{\omega=\bar\omega_{ij}}+
\frac{d\Delta \rho_{ij}(m_Q)}{d\omega}\Big|_{\omega=\bar\omega_{ij}}(\omega-\bar\omega_{ij})+
\frac{1}{2}\frac{d^2\Delta \rho_{ij}(m_Q)}{d\omega^2}\Big|_{\omega=\bar\omega_{ij}}
(\omega-\bar\omega_{ij})^2+\cdots,\label{ta}
\end{eqnarray}
where the constant $\bar\omega_{ij}$, together with the index $\alpha_{ij}$ and the coefficient 
$y_{ij}$, are fixed through the fit of the first term 
$\Delta \rho_{ij}(m_Q)|_{\omega=\bar\omega_{ij}}$ 
to the initial condition in the interval $(m_{ij},M_{ij})$ of $m_Q$. The insensitivity to the 
variable $\omega$ commands the vanishing of the first derivative in Eq.~(\ref{ta}), 
$d\Delta \rho_{ij}(m_Q)/d\omega|_{\omega=\bar\omega_{ij}}=0$, from which roots of $m_Q$ are 
solved. Furthermore, the second derivative 
$d^2\Delta \rho_{ij}(m_Q)/d\omega^2|_{\omega=\bar\omega_{ij}}$ should be minimal to maximize the 
stability window around $\bar\omega_{ij}$, in which $\Delta\rho_{ij}(m_Q)$ remains almost 
independent of $\omega$.

The threshold behaviors in Eq.~(\ref{asy}) and the initial conditions
$\Delta\rho_{ij}(m_Q)=-{\rm Im}\Pi^{\rm box}_{ij}(m_Q)$ with ${\rm Im}\Pi^{\rm box}_{ij}(m_Q)$
being defined according to Eq.~(\ref{mi}) set the initial conditions in the limits 
$m_Q\to m_{ij}$
\begin{eqnarray}
\Delta\rho_{db}(m_Q)&\sim &m_Q^2-m_b^2,\nonumber\\
\Delta\rho_{sb}(m_Q)&\sim &[m_Q^2-(m_b+m_s)^2]^{-1/2},\nonumber\\
\Delta\rho_{bb}(m_Q)&\sim &(m_Q^2-4m_b^2)^{1/2}.\label{p2}
\end{eqnarray}
The solution in Eq.~(\ref{d2}) scales in the the threshold region $m_Q\sim m_{ij}$ like
$\Delta \rho_{ij}(m_Q)\sim [m_Q^2-(m_i+m_j)^2]^{\alpha_{ij}}$ owing to the relation
$J_{\alpha}(z)\sim z^\alpha$ in the limit $z\to 0$.
Contrasting this scaling law with Eq.~(\ref{p2}), we read off the indices 
\begin{eqnarray}
\alpha_{db}=1,\;\;\;\; 
\alpha_{sb}=-1/2,\;\;\;\; \alpha_{bb}=1/2.\label{av}
\end{eqnarray}
It is clear now why we employed those modified integrands in Eq.~(\ref{mi}); the corresponding 
inputs in Eq.~(\ref{p2}) are proportional to simple powers of $m_Q^2-(m_i+m_j)^2$, so that the 
indices $\alpha_{ij}$ can be specified unambiguously. The coefficients $y_{ij}$ are
related to the boundary conditions at the high end $m_Q=M_{ij}$ of the interval 
$(m_{ij},M_{ij})$, $\Delta\rho_{ij}(M_{ij})=-{\rm Im}\Pi^{\rm box}_{ij}(M_{ij})$, which fix the 
coefficients
\begin{eqnarray}
y_{ij}=-{\rm Im}\Pi^{\rm box}_{ij}(M_{ij})\left[\left(\omega 
\sqrt{M_{ij}^2-(m_i+m_j)^2}\right)^{\alpha_{ij}} 
J_{\alpha_{ij}}\left(2\omega\sqrt{M_{ij}^2-(m_i+m_j)^2}\right)\right]^{-1}.\label{dc2}
\end{eqnarray}

\begin{figure}
\begin{center}
\includegraphics[scale=0.20]{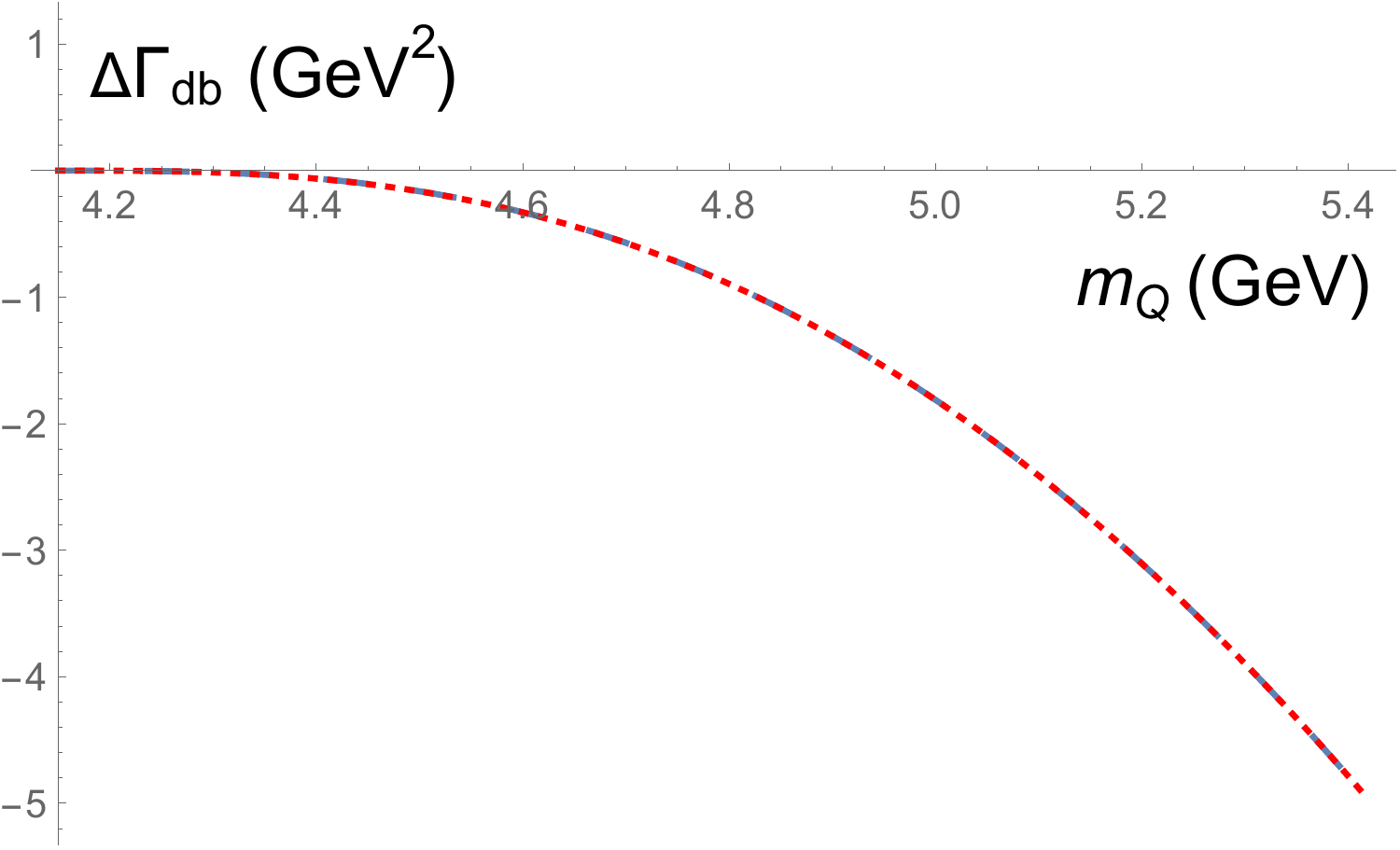}\hspace{1.0 cm} 
\includegraphics[scale=0.20]{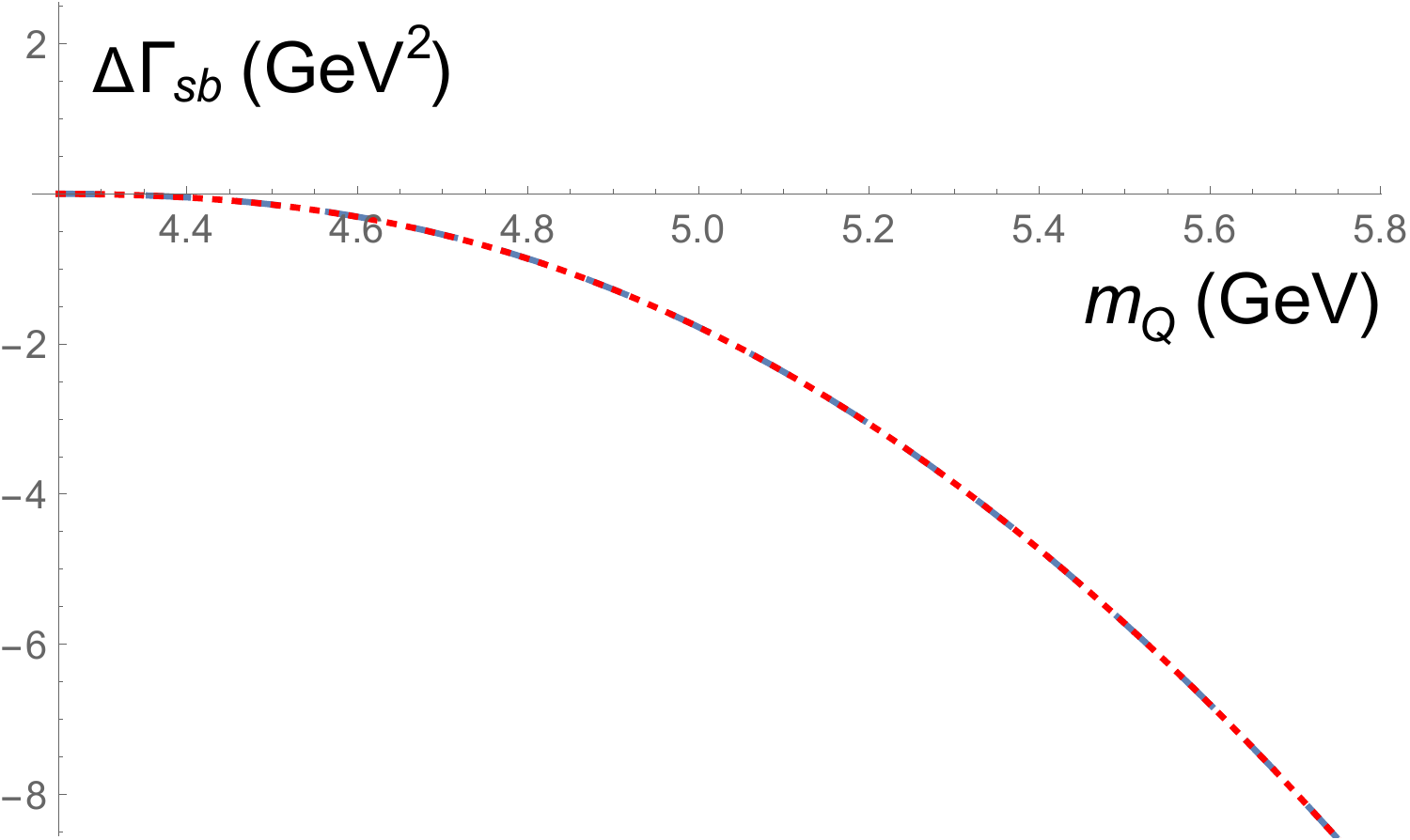}\hspace{1.0 cm} 
\includegraphics[scale=0.20]{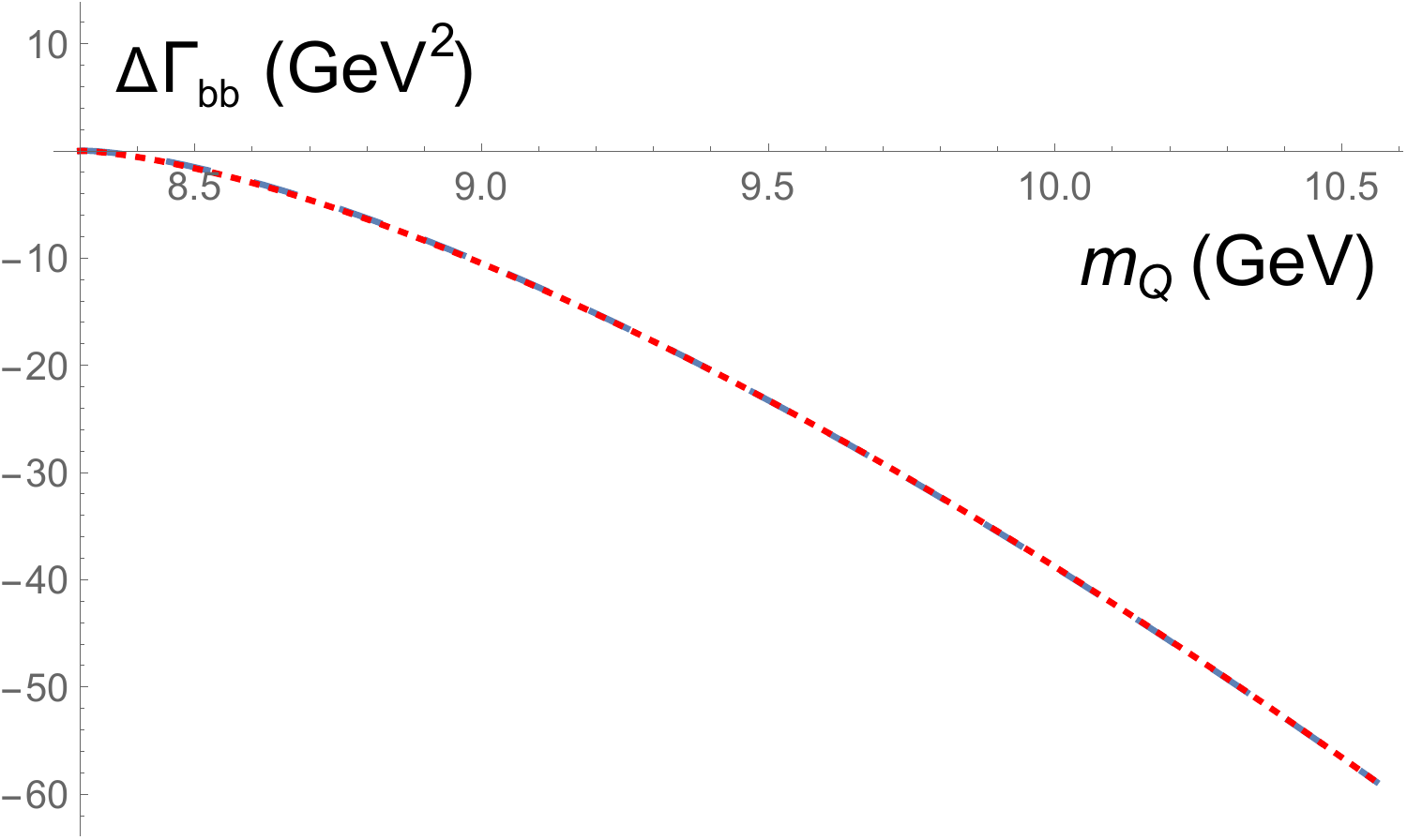}

(a) \hspace{6.0 cm} (b)\hspace{6.0 cm} (c)
\caption{\label{fig2} 
Comparison of $\Delta\Gamma_{ij}(m_Q)\equiv \Gamma_{ij}(m_Q)-\Gamma_{ij}^{\rm box}(m_Q)$ 
from the fit (dotted line) with the input $-\Gamma_{ij}^{\rm box}(m_Q)$ (dashed line) in the 
interval $(m_{ij},M_{ij})$ of $m_Q$ for (a) $ij=db$, (b) $ij=sb$, and (c) $ij=bb$.}
\end{center}
\end{figure}

The running coupling constant is given by
\begin{eqnarray}
\alpha_s(\mu)=\frac{4\pi}{\beta_0\ln(\mu^2/\Lambda_{\rm QCD}^2)}
\left[1 - \frac{\beta_1\ln\ln(\mu^2/\Lambda_{\rm QCD}^2)}
{\beta_0^2\ln(\mu^2/\Lambda_{\rm QCD}^2)}\right],
\end{eqnarray}
with the coefficients $\beta_0 = 11 - 2n_f/3$ and $\beta_1=2(51-19n_f/3)$. We take the QCD scale 
$\Lambda_{\rm QCD}=0.21$ GeV for the number of active quark flavors $n_f=5$ \cite{Zhong:2021epq}, 
and choose the renormalization scale $\mu=m_Q$ as stated before. Note that we need only the 
quark-mass inputs for the initial conditions in the interval $(m_{ij},M_{ij})$ of $m_Q$. 
Adopting the quark masses $m_s=0.1$ GeV and $m_b=4.15$ GeV in the $\overline{\rm MS}$ scheme
at the scale $\mu\sim m_b$, which are close to those from lattice calculations 
\cite{FermilabLattice:2018est}, and the pion (kaon, $B$ meson) mass $m_\pi=0.14$ GeV
($m_K=0.49$ GeV, $m_B=5.28$ GeV) \cite{PDG}, we get $\bar\omega_{db}=0.0531$ GeV$^{-1}$, 
$\bar\omega_{sb}=0.0268$ GeV$^{-1}$ $\bar\omega_{bb}=0.0128$ GeV$^{-1}$ from the best fits of 
$\Delta\rho_{ij}(m_Q)$ in Eq.~(\ref{d2}) to $-{\rm Im}\Pi^{\rm box}_{ij}(m_Q)$ in the interval 
$(m_{ij},M_{ij})$. The fit results by means of $\Delta\Gamma_{ij}(m_Q)$, which are related to 
$\Delta\rho_{ij}(m_Q)$ via Eq.~(\ref{mi}), are compared with $-\Gamma_{ij}^{\rm box}(m_Q)$ 
in the interval $(m_{ij},M_{ij})$ in Fig.~\ref{fig2}. Their perfect matches confirm that the 
approximate solutions in Eq.~(\ref{d2}) work well, and that other methods for obtaining 
$\bar\omega_{ij}$ should return similar values. For example, equating $\Delta\rho_{ij}(m_Q)$ 
and $-{\rm Im}\Pi^{\rm box}_{ij}(m_Q)$ at the midpoints $m_Q=(m_{ij}+M_{ij})/2$ leads to 
$\bar\omega_{db}=0.0503$ GeV$^{-1}$, $\bar\omega_{sb}=0.0268$ GeV$^{-1}$ and 
$\bar\omega_{bb}=0.0129$ GeV$^{-1}$, very close to those from the best fits.

\begin{figure}
\begin{center}
\includegraphics[scale=0.30]{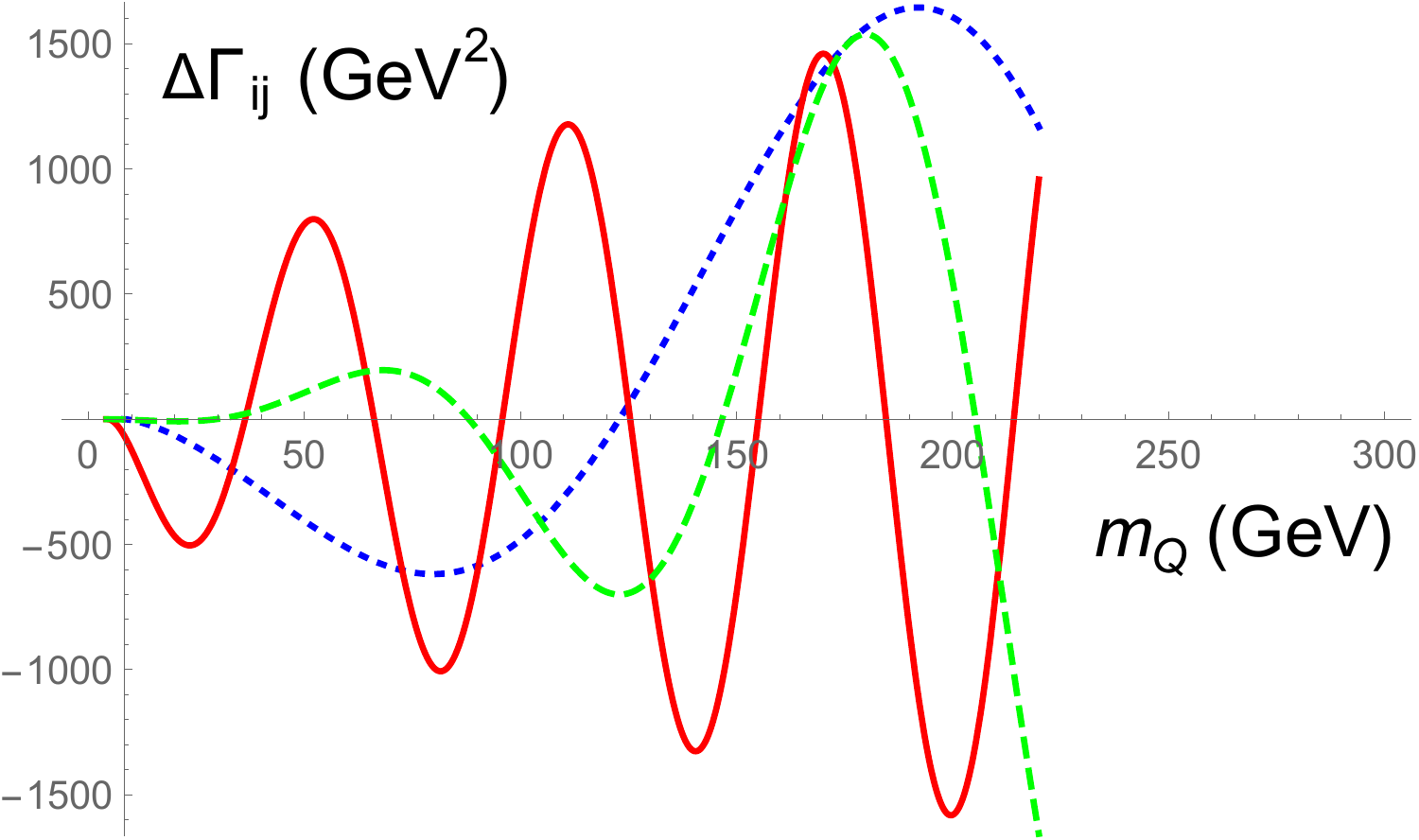}\hspace{1.0 cm} 
\includegraphics[scale=0.30]{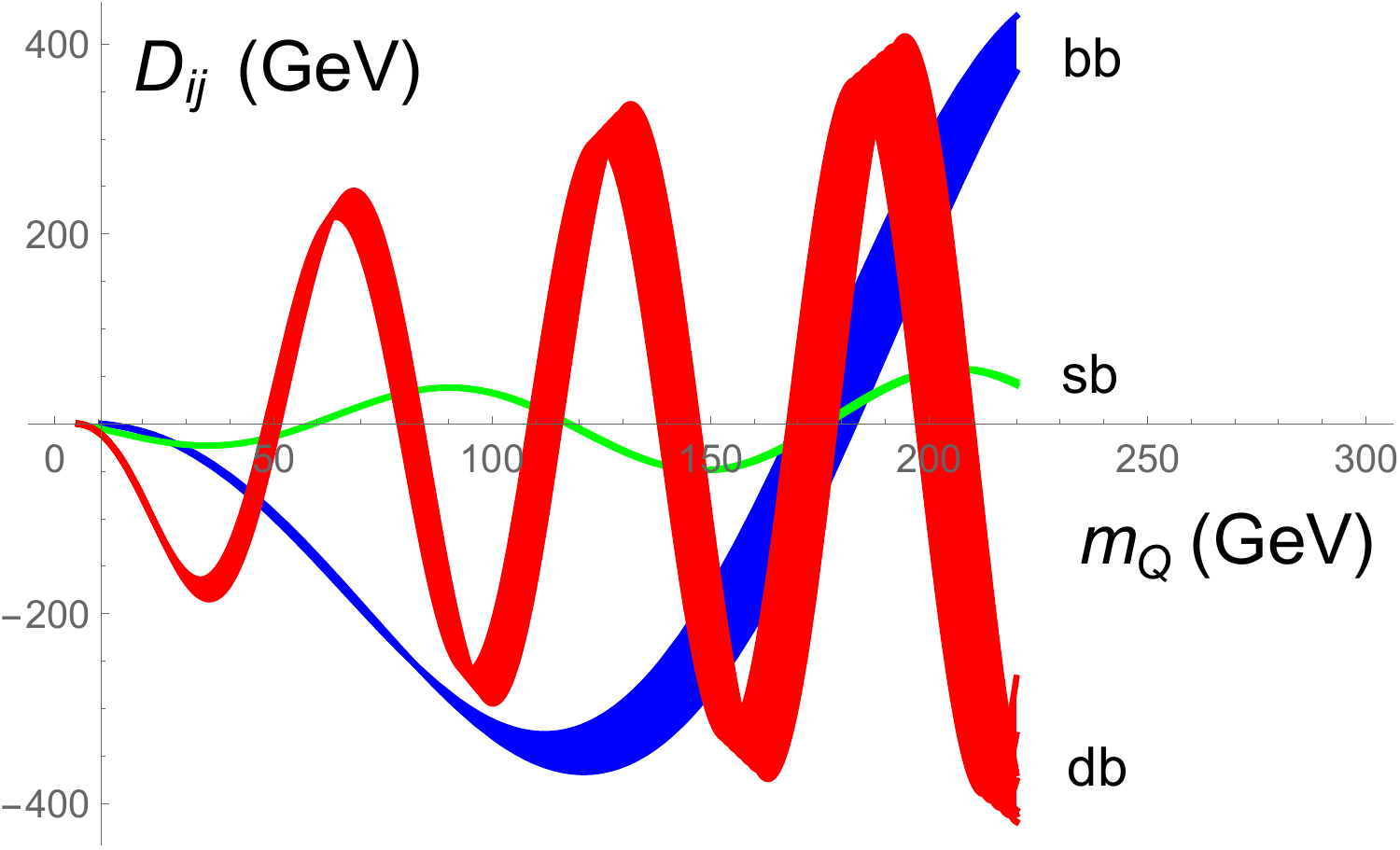}

(a) \hspace{8.0 cm} (b)

\caption{\label{fig3} 
(a) Dependencies of the solutions $\Delta\Gamma_{ij}(m_Q)$ on $m_Q$ for $ij=db$ (solid line), 
$ij=sb$ (scaled by a factor 0.02, dashed line), and $ij=bb$ (scaled by a factor 0.1, dotted line).
(b) Dependencies of the derivatives $D_{ij}(m_Q)$ on $m_Q$.}
\end{center}
\end{figure}

The unknown subtracted functions $\Delta\rho_{ij}(m_Q)$ with the above $\alpha_{ij}$, $y_{ij}$ 
and $\bar\omega_{ij}$ are displayed in 
Fig.~\ref{fig3}(a) through $\Delta\Gamma_{ij}(m_Q)$. They exhibit oscillatory behaviors in 
$m_Q$, and the first (second, third) peak of the solution for the $bb$ ($sb$, $db$) channel is 
located around $m_Q\approx 170$-195 GeV. The coincidence between the sequences of the peaks 
and of the quark generations is intriguing. The similar feature will appear again in the 
plots for the fourth generation quark masses in the next section. To evince the implication 
of the above peak overlap, we present in Fig.~\ref{fig3}(b) the dependencies of the derivatives 
$d\Delta\rho_{ij}(m_Q)/d\omega$ on $m_Q$ by 
\begin{eqnarray}
D_{ij}(m_Q)\equiv \frac{d}{d\omega}
\frac{J_{\alpha_{ij}}\left(2\omega \sqrt{m_Q^2-(m_i+m_j)^2}\right)}
{J_{\alpha_{ij}}\left(2\omega\sqrt{M_{ij}^2-(m_i+m_j)^2}\right)}\Bigg|_{\omega=\bar\omega_{ij}},
\label{dij}
\end{eqnarray}
where the factors independent of $\omega$ have been dropped for simplicity. 

The band of the $bb$ curve is induced by the variation of the bottom quark mass $m_b$ 
in the range $m_b=(4.15\pm 0.01)$ GeV with roughly 1$\sigma$ deviation from the value
$4.18^{+0.03}_{-0.02}$ GeV in \cite{PDG}. The considered error of $m_b$ is also compatible with
that obtained in Ref.~\cite{FermilabLattice:2018est}. The result for the $db$ channel is less 
sensitive to $m_b$, but depends more strongly on the methods of determining $\bar\omega_{db}$ 
as mentioned before. Namely, the band of the $db$ curve is mainly 
attributed to the latter source of uncertainties with $\bar\omega_{db}$ being lowered to 
$0.0503$ GeV$^{-1}$. The derivative $D_{sb}(m_Q)$ is stable with respect to various sources of 
uncertainties; for instance, changing the strange-quark mass $m_s$ by 10\% causes only about 
1\% effects. It is the reason why the $sb$ curve discloses a narrow band. Every 
curve in Fig.~\ref{fig3}(b) indicates the existence of multiple roots. It has been checked 
that the second derivatives are larger at higher roots \cite{Li:2023dqi}, so 
smaller roots are preferred in the viewpoint of maximizing the stability windows in $\omega$. Figure~\ref{fig3}(b) shows that the three derivatives first vanish 
simultaneously around $m_Q\approx 173$ GeV, as manifested by the intersection of the three 
curves in the interval $(170\;{\rm GeV}, 176\;{\rm GeV})$, which corresponds to the 
location of the peak overlap in Fig.~\ref{fig3}(a). To be explicit, we read off the roots
$m_Q=169.1^{+9.5}_{-1.1}$ GeV for the $db$ channel, $m_Q=176.2\pm 0.6$ GeV for the 
$sb$ channel and $m_Q=175.7^{+7.3}_{-6.3}$ GeV for the $bb$ channel in Fig.~\ref{fig3}(b).
The result of $m_Q$, as a common solution
to the considered channels, is identified as the physical top quark mass, which agrees 
well with the observed one $m_t=(172.69\pm 0.30)$ GeV \cite{PDG}.

A remark is in order. The tiny error 0.01 GeV for the input $m_b=(4.15\pm 0.01)$ GeV was 
adopted to examine the sensitivity of our predictions to the variation of the bottom quark 
mass. We emphasize that the main purpose of the present work is to predict the fourth 
generation quark masses, for which both the bottom and top quark masses are necessary 
inputs. Hence, the reproduction of the top quark mass from the given bottom quark mass in 
its allowed range is not only to validate our formalism, but to calibrate the inputs for the 
predictions. This calibration is essential owing to the sensitivity to the inputs as noticed 
above (the determination of the lighter quark masses in our formalism is more 
stable against variations of inputs \cite{Li:2023dqi}). Besides, we set 
the renormalization scale to the invariant mass $m_Q$ of the heavy quark in Eq.~(\ref{bij}), 
and stick to this choice for the consistent determination of the top quark mass and the fourth 
generation quark masses. We think that $m_b=4.15\pm 0.01$ GeV and the resultant 
$m_t=173\pm 3$ GeV, in agreement with the extractions from other known means and current data, 
serve as the appropriate inputs. Note that only the outcome from the $bb$ channel, 
which involves two bottom quarks in the intermediate states, is sensitive to the input of $m_b$.
Therefore, a resolution to the aforementioned sensitivity that one can make
is to discard the $bb$ channel, and to consider simply the $db$ and $sb$ channels. The 
simultaneous vanishing of their derivatives in Eq.~(\ref{dij}) is sufficient for deriving 
a stable and definite top quark mass. 

%We do not intend to enlarge their uncertainties by increasing the error of the bottom quark 
%mass or varying the renormalization scale.

\section{FOURTH GENERATION QUARK MASSES}

After verifying that the dispersive analysis produces the correct top quark mass, we extend it 
to the predictions of the fourth generation quark masses, starting with the $b'$ one. Consider 
the box diagrams for the mixing of the neutral states $Q\bar d$ and $\bar Qd$, and 
construct the associated dispersion relations. The intermediate channels, which contribute to 
the imaginary pieces of the box diagrams, contain not only those from on-shell quarks $ut$, 
$ct$ and $tt$ described by Eq.~(\ref{bij}), but those from on-shell $W$ bosons. Since these 
channels can be differentiated experimentally, we can focus on the former for our purpose. The 
necessary power corrections proportional to the differences between the quark-level 
thresholds $m_{ij}$ and the physical thresholds $M_{ij}$ further select the $ut$ channel with  
$m_{ut}=m_t$ ($m_u=0$) and $M_{ut}=m_\pi+m_t$, and the $ct$ channel with $m_{ct}=m_c+m_t$ and 
$M_{ct}=m_D+m_t$, $m_D$ being the $D$ meson mass. Note that the second term in the curly 
brackets of Eq.~(\ref{bij}) becomes more important in the present case owing to the large ratio 
$(m_i^2+m_j^2)/m_W^2\approx m_t^2/m_W^2$. Equation~(\ref{bij}) behaves approximately 
in the threshold regions with $m_Q\sim m_{ij}$ like
\begin{eqnarray}
\Gamma^{\rm box}_{ut}(m_Q)&\sim &\frac{(m_Q^2-m_t^2)^2}{m_Q^2},\nonumber\\
\Gamma^{\rm box}_{ct}(m_Q)&\sim&\frac{m_Q^2-m_t^2-m_c^2}
{m_Q^2}\sqrt{[m_Q^2-(m_t+m_c)^2][m_Q^2-(m_t-m_c)^2]}.\label{as2}
\end{eqnarray}
Because of $m_c\ll m_t$, the terms $(m_t-m_c)^2$ and $m_t^2+m_c^2$, which are not very 
distinct from $(m_t+m_c)^2$, have stayed in the second line of Eq.~(\ref{as2}).

Motivated by the above threshold behaviors, we choose the integrands for the 
dispersion integrals in Eq.~(\ref{it0}) as
\begin{eqnarray}
{\rm Im}\Pi_{ut}(m)&= &\frac{m^2\Gamma_{ut}(m)}{m^2-m_t^2},\nonumber\\
{\rm Im}\Pi_{ct}(m)&= &\frac{m^2\Gamma_{ct}(m)}
{[m^2-(m_t+m_c)^2](m_Q^2-m_t^2-m_c^2)\sqrt{m^2-(m_t-m_c)^2}}.\label{mi2}
\end{eqnarray}
Similarly, the above integrands with 
powers of $m$ in the numerators suppress the residues from the poles at $m=\pm(m_i+m_j)$
and $m=\pm\sqrt{m_i^2+m_j^2}$ in the low $m$ region, compared to the ones from $m=\pm m_Q$ at 
large $m_Q$. Our contour for $\Pi_{ct}(m)$ crosses the real axis between $m=-(m_t+m_c)$ and 
$m=-\sqrt{m_t^2+m_c^2}$ and between $m=\sqrt{m_t^2+m_c^2}$ and $m=m_t+m_c$, and runs along the 
real axis marked by $m<-(m_t+m_c)$ and $m>m_t+m_c$. Therefore, the branch cut associated with 
the factor $\sqrt{m^2-(m_t-m_c)^2}$ in $\Pi_{ct}(m)$ does not contribute. The solutions for the 
unknown subtracted functions $\Delta \rho_{ij}(m_Q)$ and the coefficients $y_{ij}$ have the 
same forms as Eqs.~(\ref{d2}) and (\ref{dc2}), respectively. The initial conditions near the 
thresholds $m_Q\sim m_{ij}$ are given by  
\begin{eqnarray}
\Delta\rho_{ut}(m_Q)&\sim &m_Q^2-m_t^2,\nonumber\\
\Delta\rho_{ct}(m_Q)&\sim &[m_Q^2-(m_t+m_c)^2]^{-1/2},\label{p22}
\end{eqnarray}
which assign the indices
\begin{eqnarray}
\alpha_{ut}=1,\;\;\;\; \alpha_{ct}=-1/2.\label{av2}
\end{eqnarray}

For the numerical study, we take the QCD scale $\Lambda_{\rm QCD}^{(6)}=0.11$ GeV for the 
number of active quark flavors $n_f=6$ according to \cite{Deur:2016tte}
\begin{eqnarray}
\Lambda_{\rm QCD}^{(n_f)}=\Lambda_{\rm QCD}^{(n_f-1)}
\left[\frac{\Lambda_{\rm QCD}^{(n_f-1)}}{m_t}\right]^{2/(3\beta_0)},\label{Ln}
\end{eqnarray}
with $m_t=173$ GeV and $\Lambda_{\rm QCD}^{(5)}=0.21$ GeV \cite{Zhong:2021epq}.   
The behaviors of the box-diagram contributions $\Gamma^{\rm box}_{ij}(m_Q)$ in the interval 
$(m_{ij},M_{ij})$ of $m_Q$ matter in solving the dispersion relations. In view of the high 
top-quark mass, the renormalization-group (RG) evolution of the charm-quark mass to a scale of 
$O(m_t)$ needs to be taken into account. This RG effect is minor in the previous section, since
$m_b$ does not deviate much from the range $\mu\approx 1$-2 GeV, in which the strange-quark mass is 
defined. We have at $\mu=m_t$ \cite{Zhong:2021epq}
\begin{eqnarray}
m_c(m_t)=m_c(m_c)\left[\frac{\alpha_s(m_t)}{\alpha_s(m_c)}\right]^{4/\beta_0}
\approx 0.7\;{\rm GeV},\label{rg}
\end{eqnarray}
for $m_c(m_c)=1.35$ GeV \cite{Li:2023dqi}. The inputs of the pion mass $m_\pi=0.14$ GeV and the 
$D$-meson mass $m_D=1.87$ GeV \cite{PDG} then yield $\bar\omega_{ut}=0.00326$ GeV$^{-1}$ 
and $\bar\omega_{ct}=0.00176$ GeV$^{-1}$ from the best fits of Eq.~(\ref{d2}) to 
$-{\rm Im}\Pi^{\rm box}_{ij}(m_Q)$ in the interval $(m_{ij},M_{ij})$ of $m_Q$, which is 
proportional to Eq.~(\ref{bij}). Equating 
$\Delta\rho_{ij}(m_Q)$ and $-\Pi^{\rm box}_{ij}(m_Q)$ at the midpoints $m_Q=(m_{ij}+M_{ij})/2$ 
generates $\bar\omega_{ut}=0.00326$ GeV$^{-1}$ and $\bar\omega_{ct}=0.00175$ GeV$^{-1}$, 
almost identical to the values from the best fits. This consistency supports the goodness
of our solutions.

\begin{figure}
\begin{center}
\includegraphics[scale=0.30]{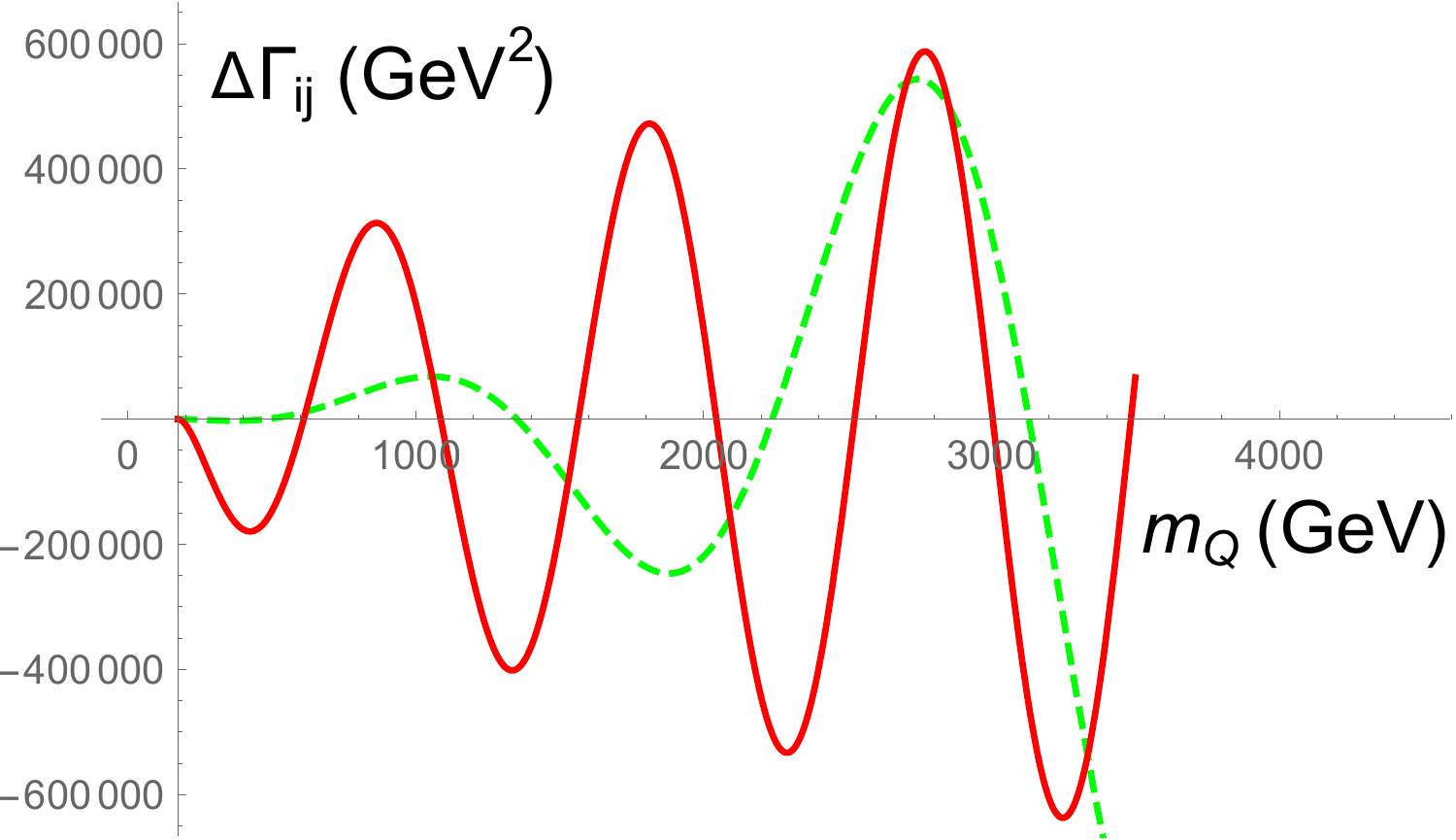}\hspace{1.0 cm}
\includegraphics[scale=0.30]{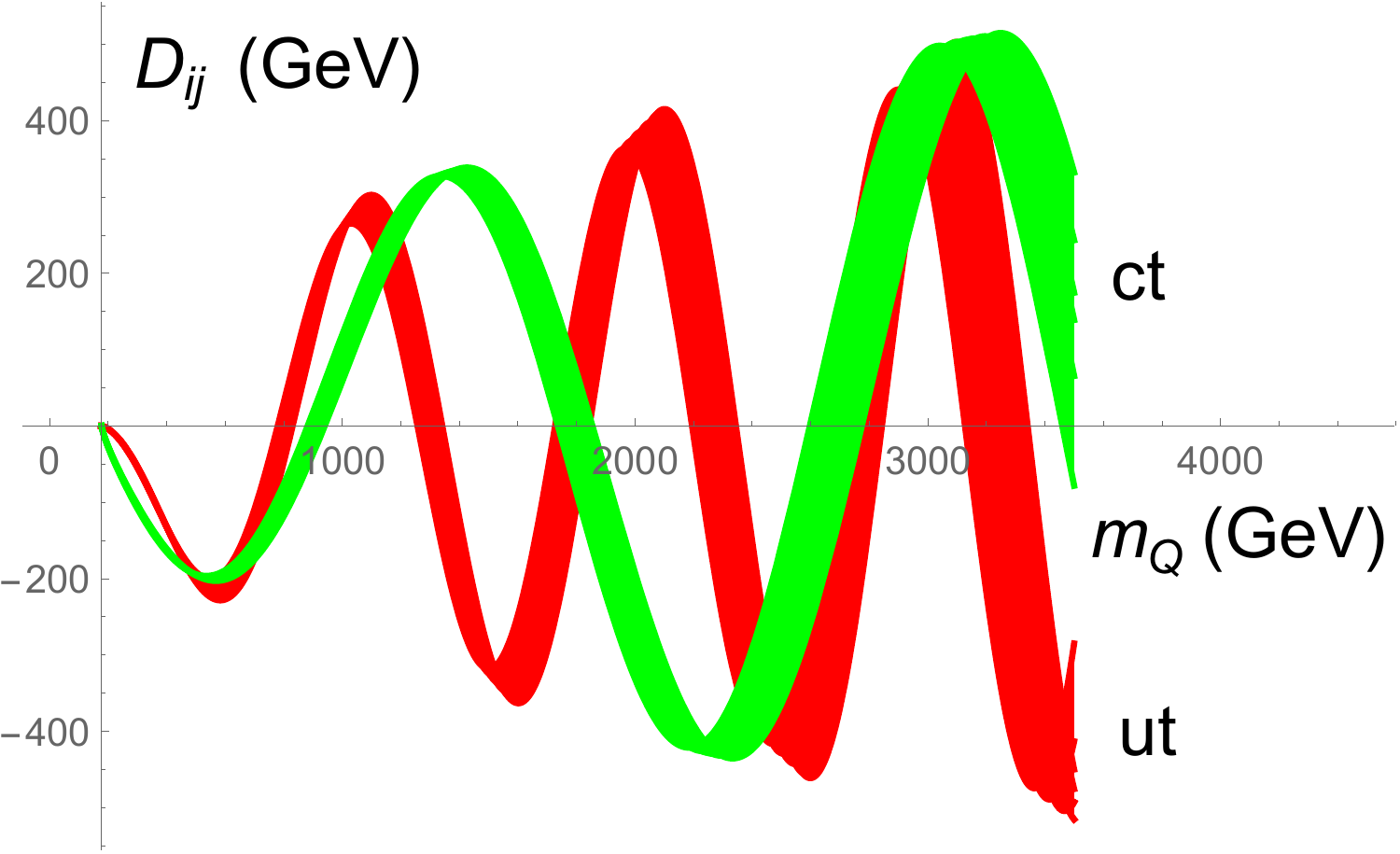}

(a) \hspace{8.0 cm} (b)

\caption{\label{fig4} 
(a) Dependencies of the solutions $\Delta\Gamma_{ij}(m_Q)$ on $m_Q$ for $ij=ut$ (solid line) and 
$ij=ct$ (scaled by a factor 0.02, dashed line).
(b) Dependencies of the derivatives $D_{ij}(m_Q)$ on $m_Q$.
The curve for $ij=ut$ has been scaled by a factor 0.01.}
\end{center}
\end{figure}

The dependencies of the unknown subtracted functions $\Delta\rho_{ij}(m_Q)$ on $m_Q$ from solving 
the dispersion relations are presented in Fig.~\ref{fig4}(a) by means of 
$\Delta\Gamma_{ij}(m_Q)$. We have confirmed the excellent matches between $\Delta\Gamma_{ij}(m_Q)$ 
form the fits and the initial conditions $-\Gamma_{ij}^{\rm box}(m_Q)$ in the interval 
$(m_{ij},M_{ij})$ of $m_Q$, which will not be shown here. The feature noticed before hints 
that the second (third) peak of the curve for the $ct$ ($ut$) channel should be located at roughly 
the same $m_Q$. Figure~\ref{fig4}(a), with the overlap of peaks around $m_Q\approx 2.7$ TeV, 
corroborates this expectation. The corresponding derivatives in Eq.~(\ref{dij}) 
as functions of $m_Q$ are drawn in Fig.~\ref{fig4}(b). Similarly, our results for the $ct$ 
channel are insensitive to the variation of $m_c$: 10\% changes of $m_c$ stimulates only about 
1\% effects on the outcome of the fourth generation quark mass $m_{b'}$. The uncertainties 
from different ways of fixing $\bar\omega_{ij}$ are negligible as investigated above. Hence, we 
consider only the uncertainties from the variation of the top-quark mass within $m_t=(173\pm 3)$ 
GeV attained in the previous section, which are reflected by the bands of the curves. It is 
found that the two derivatives first vanish simultaneously around $m_Q\approx 2.7$ TeV, 
coinciding with the location of the peak overlap in Fig.~\ref{fig4}(a). That is, a common 
solution $m_{b'}=(2.7\pm 0.1)$ TeV, as inferred from Fig.~\ref{fig4}(b), exists for the two 
considered channels.

\begin{figure}
\begin{center}
\includegraphics[scale=0.30]{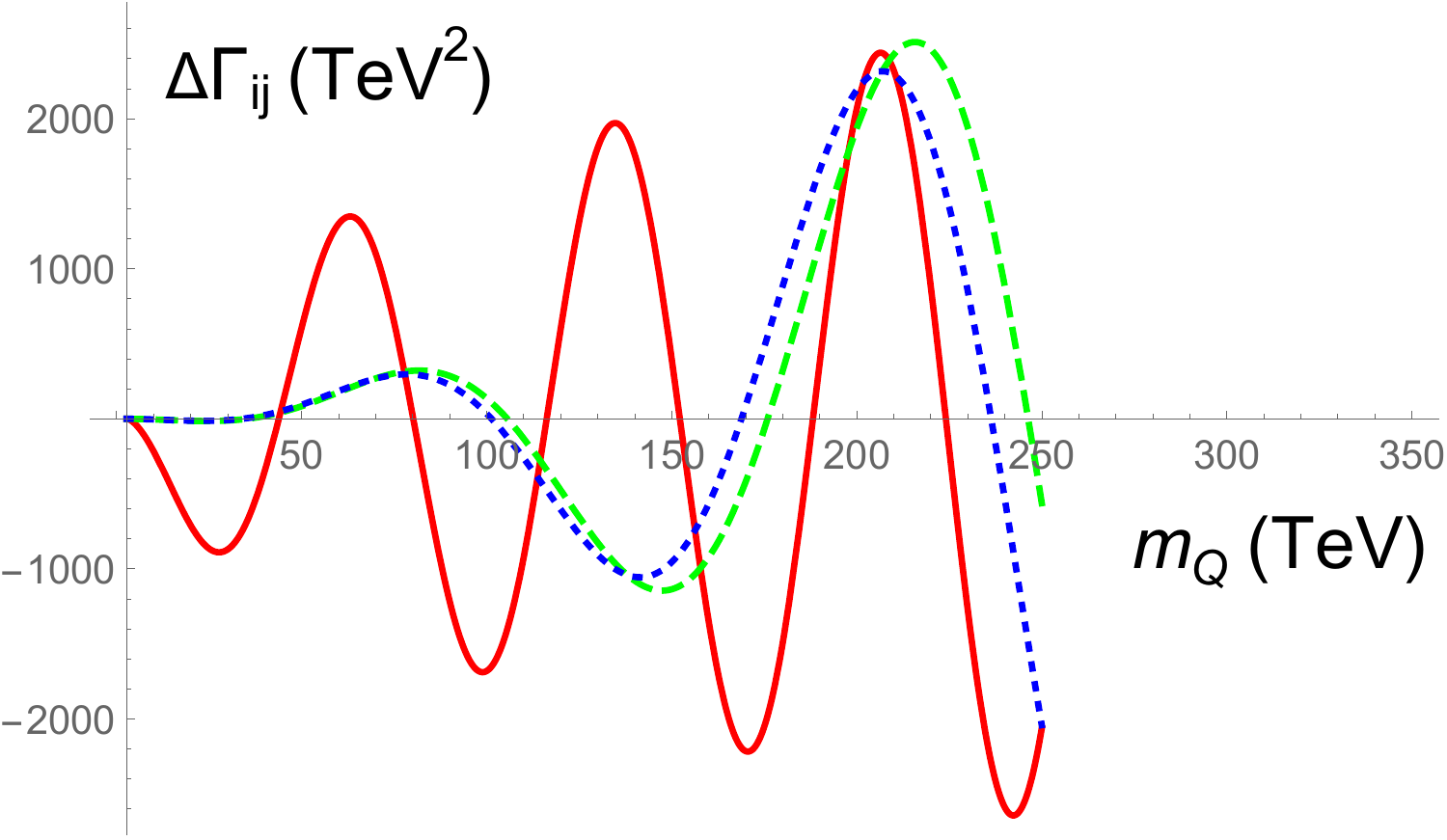}\hspace{1.0 cm}
\includegraphics[scale=0.30]{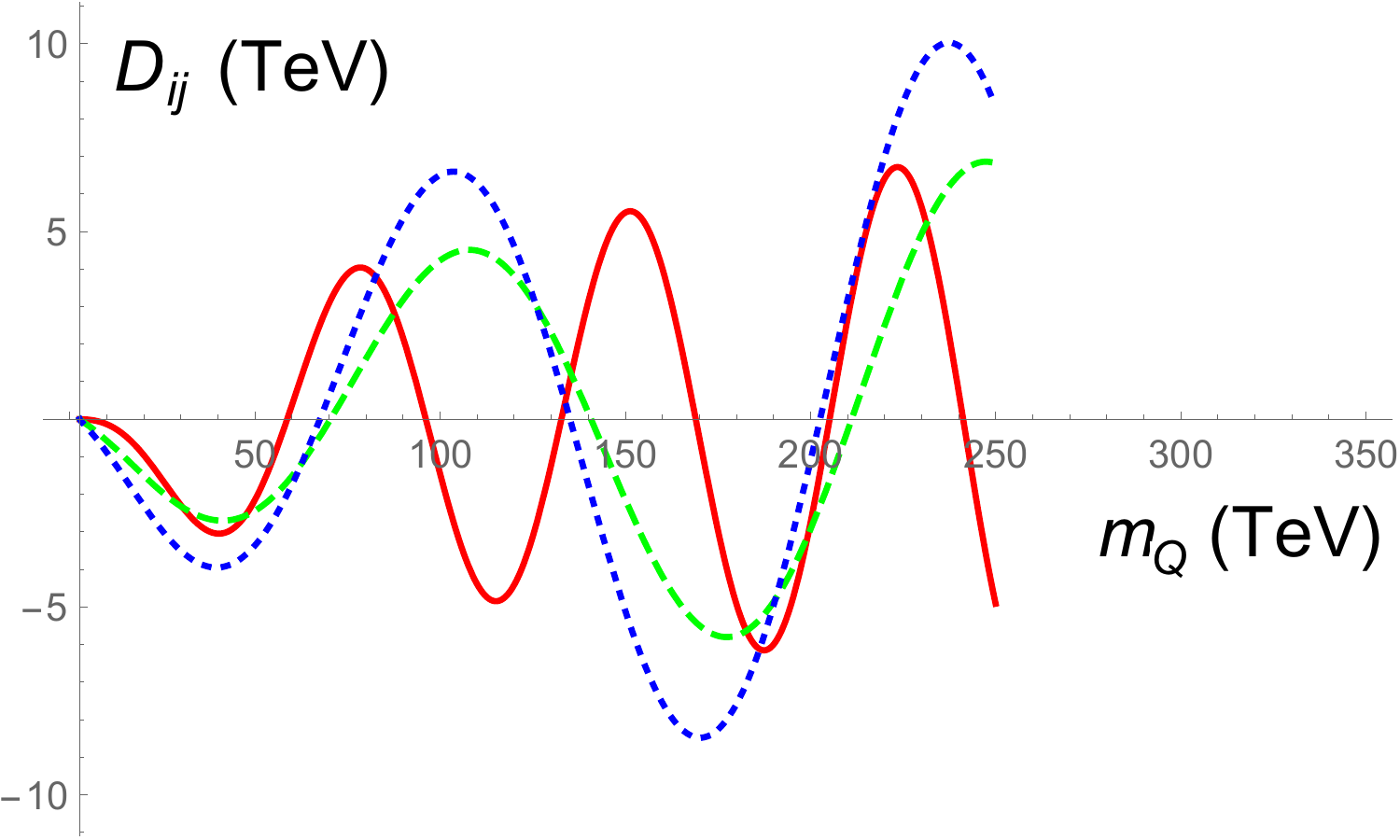}

(a) \hspace{8.0 cm} (b)

\caption{\label{fig5} 
(a) Dependencies of the solutions $\Delta\Gamma_{ij}(m_Q)$ on $m_Q$ for $ij=db'$ (solid line), 
$ij=sb'$ (scaled by a factor 0.02, dashed line), and $ij=bb'$ (scaled by a factor 0.02, dotted line).
(b) Dependencies of the derivatives $D_{ij}(m_Q)$ on $m_Q$
for $ij=db'$ (scaled by a factor $10^{-4}$, solid line), 
$ij=sb'$ (dashed line), and $ij=bb'$ (dotted line).}
\end{center}
\end{figure}

The prediction of the fourth generation quark mass $m_{t'}$ proceeds in exactly the same 
manner. The box diagrams governing the mixing of the neutral states $Q\bar u$ and $\bar Q u$
involve the intermediate $db'$, $sb'$ and $bb'$ channels, which are associated with
the quark-level thresholds $m_{db'}=m_{b'}$ ($m_d=0$), $m_{sb'}=m_s+m_{b'}$ and 
$m_{bb'}=m_b+m_{b'}$, and the physical thresholds $M_{db'}=m_\pi+m_{b'}$, $M_{sb'}=m_K+m_{b'}$ 
and $M_{bb'}=m_B+m_{b'}$, respectively. Since a top quark does not form a 
hadronic bound state, we do not expect that a $b'$ quark will, and keep the quark mass 
$m_{b'}$ in the hadronic thresholds. Certainly, this is an assumption owing to the
the uncertain $4\times 4$ CKM matrix element $V_{tb'}$.
The second term in the curly brackets of Eq.~(\ref{bij}) 
dominates because of the large ratio $(m_i^2+m_j^2)/m_W^2\approx m_{b'}^2/m_W^2$. The behaviors 
of Eq.~(\ref{bij}) in the threshold regions with $m_Q\sim m_{ij}$ are approximated by
Eq.~(\ref{as2}), with the first line for the $db'$ channel and the second line for the $sb'$ 
and $bb'$ channels. The appropriate replacements of the masses $m_{c,t}$ by $m_{s,b,b'}$ are
understood. The modified integrands for the dispersion integrals in Eq.~(\ref{it0}) and 
their expressions near the thresholds $m_Q\sim m_{ij}$ follow Eqs.~(\ref{mi2}) and (\ref{p22}), 
respectively, also with the first lines for the $db'$ channel and the second lines for the 
$sb'$ and $bb'$ channels. We then acquire the indices
\begin{eqnarray}
\alpha_{db'}=1,\;\;\;\; \alpha_{sb'}=\alpha_{bb'}=-1/2.\label{av3}
\end{eqnarray}

The QCD scale takes the value $\Lambda_{\rm QCD}^{(7)}=0.04$ GeV for $n_f=7$ according to 
Eq.~(\ref{Ln}) but with $m_{b'}$ being substituted for $m_t$. The RG effects on the quark 
masses are included via Eq.~(\ref{rg}), which give $m_s\approx 0.07$ GeV and 
$m_b\approx 3.2$ GeV at the scale $\mu=m_{b'}$. Inputting the same masses $m_\pi$, $m_K$, 
$m_B$ and $m_{b'}=2.7$ TeV, we get $\bar\omega_{db'}=0.0438$ TeV$^{-1}$, 
$\bar\omega_{sb'}=0.0223$ TeV$^{-1}$, and $\bar\omega_{bb'}=0.0233$ TeV$^{-1}$ from the 
best fits of Eq.~(\ref{d2}) to $-{\rm Im}\Pi^{\rm box}_{ij}(m_Q)$ in the interval 
$(m_{ij},M_{ij})$ of $m_Q$. Figure~\ref{fig5}(a) collects the solutions $\Delta\Gamma_{ij}(m_Q)$ 
as functions of $m_Q$, where the curves for the $bb'$ and $sb'$ channels are close in shape, 
and their second peaks overlap with the third peak for the $db'$ channel around 
$m_Q\approx 200$ TeV. The $bb'$ and $sb'$ channels share the identical formula characterized 
by the same indices $\alpha_{sb'}=\alpha_{bb'}=-1/2$. Moreover, the difference between $m_s$ 
and $m_b$ (also between $m_K$ and $m_B$) is minor relative to the high $m_{b'}$, so that these 
two solutions behave similarly. Hence, there are only two categories of solutions in the $t'$ 
case, and the overlap takes place between the second and third peaks.

The dependencies of the derivatives $D_{ij}(m_Q)$ on $m_Q$, defined in Eq.~(\ref{dij}),  
are displayed in Fig.~\ref{fig5}(b) for $\omega=\bar\omega_{ij}$. The three derivatives 
first vanish simultaneously around $m_Q\approx 200$ TeV, which coincides with the 
aforementioned peak locations. It is sure that a common root for the fourth generation quark 
mass $m_{t'}$ exists. Since the value of $m_{t'}$ is obviously beyond the current 
and future reach of new particle searches, we do not bother to include theoretical 
uncertainties with the prediction. One may wonder whether $m_{t'}\approx 200$ TeV 
violates the unitarity limit signified by the large Yukawa coupling. However, bound states 
would be formed in this case, such that physical degrees of freedom change, 
and the high Yukawa coupling is not an issue. This subject will be elaborated in the 
next section. It is not unexpected that a $t'$ quark is so heavy, viewing that a $c$ quark 
is 13 times heavier than an $s$ quark, and a $t$ quark is about 40 times heavier than a $b$ 
quark. Here a $t'$ quark is about 70 times heavier than a $b'$ quark.

\section{$\bar b'b'$ BOUND STATES}

%In the trianlge quark loop, the contributions from the $l\to\infty$ region cancel
% between the two terms $4l^\mu l^\nu-g^{\mu\nu} l^2$. The contributions from the 
%$l\sim O(m_Q)$ region cancels between the terms $m_Q(l^\mu l^\nu-m_Q^2)$. Hence,
%the loop integral for the Higgs production us diminated by the region with $l<m_Q$.
%Precisely, it is dominated by the $l^2\sim m_H^2$ region. Then only the term 
%$p_1^\mu p_2^\nu$ survives. The measure scales like $m_Q^4$ in the region of $l\sim m_Q$, 
%the numerator of the integrand scales like $m_Q p_1^\mu p_2^\nu \sim m_Qm_H^2$, the 
%Yukawa coupling scales like $m_Q/m_H$, and the denominator scales like 
%$m_Q^6$. Combining the above scaling behaviors, we get the Higgs production amplitude 
%$m_Q^4 m_Qm_H^2 (m_Q/m_H)/m_Q^6\sim m_H$. The region with $l$ above $m_Q$ contributes
%to the loop integral at the power $\sim m_H^2/m_Q.
%It explains why the cross sections scals with $m_Q$.

As remarked in the Introduction, the sequential fourth generation model is disfavored  by the 
data of Higgs boson production via gluon fusion and decay into photon pairs \cite{Chen:2012wz}. Nevertheless, it has been known \cite{Hung:2009hy} that the fourth generation 
quarks, whose mass $m_Q$ meets the criterion $K_Q=m_Q^3/(4\pi v^2 m_H)>1.68$, with the vacuum
expectation value $v=246$ GeV, form bound states in a Yukawa potential. The binding energy 
for the $\bar QQ$ ground state with the masses $m_Q^{\ast}\approx 1.26$ TeV and 
$m_H^{\ast}\approx 1.45$ TeV at the fixed point of the RG evolution in this 
model was found to be $-4.9$ GeV. The fixed point depends on the initial values 
of the quark masses at the electroweak scale of $O(100)$ GeV: the larger the initial values, the 
lower the fixed point is. The $b'$ quark mass $m_{b'}=2.7$ TeV predicted in the previous section, 
greater than the fixed-point value 1.26 TeV, satisfies the criterion $K_Q>1.68$ 
definitely. The binding energy for the $\bar b'b'$ bound state ought to be higher.
We will demonstrate that the new scalars $S$ formed by $\bar b'b'$, with
tiny couplings to a Higgs boson, escape the current experimental constraints.
It is then worthwhile to keep searching for a superheavy $b'$ quark at future colliders
\cite{Hou:2012df}.

Once the bound state of mass at TeV scale is formed, the gluon fusion process involving 
internal $b'$ quarks at the low scale $m_H$ should be analyzed in an effective theory with 
different physical degrees of freedom. In other words, one has to regard the process as gluon 
fusion into the scalar $S$, followed by production of a Higgs boson through a coupling between 
them. The order of magnitude of the corresponding amplitude is assessed below. First, the 
gluon fusion into $S$ is proportional to $\sqrt{s}g_{ggS}$, where the invariant mass
$\sqrt{s}$ of $S$ takes into account the dimension of the effective operator $A^\mu A^\nu S$,
$A^\mu$ being a gluon field, and $g_{ggS}$ is a dimensionless effective coupling.
The scalar $S$ then propagates according to a Breit-Wigner factor $1/(s-m_S^2-i\sqrt{s}\Gamma_S)$, 
where $\Gamma_S$ denotes the width of $S$. At last, $S$ transforms into a Higgs boson 
$H$ with the magnitude being described by $sg_{SH}$, 
where $g_{SH}$ is a dimensionless effective coupling. The total amplitude is thus written, in the 
effective approach, as
\begin{eqnarray}
{\cal M}\sim \frac{\sqrt{s}^3g_{ggS}g_{SH}}{s-m_S^2-i\sqrt{s}\Gamma_S},\label{amp}
\end{eqnarray}
where factors irrelevant to our reasoning have been suppressed.

Properties of heavy quarkonium states, like $\bar b'b'$, in a Yuakawa potential 
\begin{eqnarray}
V(r)=-\alpha_Y\frac{e^{-m_H^{\ast} r}}{r},
\end{eqnarray}
of the strength $\alpha_Y=m_{b'}^2/(4\pi v^2)$, have been explored intensively in the 
literature (for a recent reference, see \cite{Napsuciale:2021qtw}). With the involved 
superheavy quark mass scale, we have adopted the fixed-point Higgs boson mass $m_H^{\ast}$ in the 
exponential. Note that the number of bound states is finite for a Yukawa potential, 
distinct from the case for a Coulomb potential which allows infinitely many bound-state 
solutions. It turns out that only the states characterized by $(n,l)=(1,0)$, $(2,0)$, $(2,1)$, 
$(3,0)$ and $(3,1)$ are bounded, $n$ ($l$) being the principal (angular momentum) quantum number. The 
states labeled by $(n,l)=(3,2)$ or higher quantum numbers are not bounded. The ground state 
with $(n,l)=(1,0)$, being either a pseudoscalar 
or a vector, is expected to have a negligible coupling to a Higgs boson. It is easy 
to read off the value $\epsilon_{10}=-0.75$ of this state from Fig.~1 in 
\cite{Napsuciale:2021qtw}, i.e., the binding energy 
$E_{10}\equiv \alpha_Y^2m_{b'}\epsilon_{10}/4\approx -41$ TeV, for the parameter 
$1/(m_H^{\ast}a_0)=8.9$, $a_0\equiv 2/(\alpha_Y m_{b'})$ being the Bohr radius. It is apparent that
this deep ground state has revealed the nonrelativistic Thomas collapse \cite{TH}, and 
calls for a relativistic treatment \cite{Enkhbat:2011vp,Ikhdair:2012zz}.

To examine the coupling to a Higgs boson, we concentrate on the $P$-wave scalar states with $l=1$, 
and deduce the value $\epsilon_{21}=-0.08$ for the $(n,l)=(2,1)$ state from Fig.~2 in 
\cite{Napsuciale:2021qtw}, i.e., the binding energy 
$E_{21}\equiv \alpha_Y^2m_{b'}\epsilon_{21}/4\approx -4.96$ TeV. We suspect 
that this deep bound state also suffers from the Thomas collapse, but continue our 
order-of-magnitude estimate for completeness. 
Figure~5 in \cite{Napsuciale:2021qtw} provides the first derivative of the corresponding wave 
function at the origin
\begin{eqnarray}
32\pi a_0^5|\psi^\prime_{21}(0)|^2\approx 0.7,
\end{eqnarray}
for the parameter $\delta=m_H^{\ast}a_0=0.11$. The width $\Gamma_S$ is then approximated by 
the $S\to gg$ decay width as in the heavy quarkonium case \cite{Lansberg:2009xh},
\begin{eqnarray}
\Gamma_S=48\alpha_S^2(2m_{b'})\frac{|R'_{21}(0)|}{m_S}\approx 570\;\;{\rm GeV},\label{ga2}
\end{eqnarray}
where the strong coupling has been evaluated at twice of the $b'$ quark mass, 
$R'_{nl}(0)=\sqrt{4\pi/3}\psi^\prime_{nl}(0)$ is the derivative of the radial wave function 
at the origin \cite{Napsuciale:2021qtw}, and the scalar has the mass  
$m_S=2m_{b'}+E_{21}\approx 440$ GeV. The width in Eq.~(\ref{ga2}) is larger 
than the scalar mass, signaling another warning to the consistency of this state.

To pin down the product of the effective couplings $g_{ggS}g_{SH}$, we match the amplitude 
in Eq.~(\ref{amp}) to the one in the fundamental theory by considering the production of a 
fictitious Higgs boson with mass squared $s\approx m_S^2$. The involved scale is so high, 
that the evaluation in the fundamental theory \cite{GGM,Spira:1995rr} based on the direct 
annihilation of the $\bar b'b'$ quark pair ought to yield a result the same as in the effective 
approach. We identify the part of the amplitude, which approaches 3/2 in the lowest-order 
expression from the fundamental theory \cite{GGM,Spira:1995rr},
\begin{eqnarray}
\left|\frac{v}{s}\frac{\sqrt{s}^3g_{ggS}g_{SH}}{s-m_S^2-i\sqrt{s}\Gamma_S}\right|^2\approx 
\left(\frac{v g_{ggS}g_{SH}}{\Gamma_S}\right)^2\approx \left(\frac{3}{2}\right)^2.\label{eff}
\end{eqnarray}
The factor $s-m_S^2\ll \sqrt{s}\Gamma_S$ has been ignored in the denominator for $s\approx m_S^2$
on the right-hand side of the first equal sign. Equation~(\ref{eff}) implies 
$g_{ggS}g_{SH}=(2/3)\Gamma_S/v$ obviously. We then obtain, by extrapolating Eq.~(\ref{amp}) to 
$s=m_H^2$, the suppression factor on the $S$ contribution relative to the top-quark one in 
the SM, 
\begin{eqnarray}
\left|\frac{v}{s}\frac{\sqrt{s}^3g_{ggS}g_{SH}}{s-m_S^2-i\sqrt{s}\Gamma_S}\right|^2
\approx \left(\frac{2}{3}\frac{m_H\Gamma_S}{m_S^2}\right)^2\approx  6.2\%.\label{per}
\end{eqnarray}
The above result also suggests that the $S$ contribution decreases like $m_S^{-4}$.

We repeat the discussion for the $(n,l)=(3,1)$ state, whose binding energy and
the first derivative of the corresponding wave function at the origin read
\begin{eqnarray}
E_{31}\equiv \frac{1}{4}\alpha_Y^2m_{b'}\epsilon_{31}\approx -124\;\;{\rm GeV},\;\;\;\;
\frac{729}{8}\pi a_0^5|\psi^\prime_{31}(0)|^2\approx 0.2,
\end{eqnarray}
with $\epsilon_{31}=-0.002$ according to Figs.~2 
and 5 in \cite{Napsuciale:2021qtw}, respectively. The width $\Gamma_S$ in Eq.~(\ref{ga2}) is 
given, for this state, by
\begin{eqnarray}
\Gamma_S=48\alpha_S^2(2m_{b'})\frac{|R'_{31}(0)|}{m_S}\approx 694\;\;{\rm GeV},\label{ga3}
\end{eqnarray}
with $m_S=2m_{b'}+E_{31}\approx 5.28$ TeV. The similar matching procedure leads to the
diminishing suppression factor 
\begin{eqnarray}
\left(\frac{2}{3}\frac{m_H\Gamma_S}{m_S^2}\right)^2\approx 4.3\times 10^{-6},
\end{eqnarray}
on the $S$ contribution to the Higgs boson production via gluon fusion in the SM.

We confront the above estimates with those from the relativistic calculation
\cite{Ikhdair:2012zz}, whose Eq.~(28) indeed allows only the bound-state solutions characterized
by $n=1$, 2 and 3. Because of their crude approximation, the states labeled by the same $n$ 
but different $l$ are degenerate in eigenenergies. We take the positive eigenenergy $E_{n}$
from Eq.~(28) of \cite{Ikhdair:2012zz}, extract the binding energy $E_n^b=E_{n}-m_{b'}/2$ with
$m_{b'}/2$ being the reduced mass of the $\bar b'b'$ system, and derive the bound-state mass
$m_S=2m_{b'}+E_n^b$. It is trivial to get the ground-state mass $3.23$ TeV, the mass of
the first excited state $4.45$ TeV for $n=2$ and the mass of the second excited state
$\lesssim 5.40$ TeV for $n=3$. The last value, differing from the nonrelativistic one 
5.28 TeV by only 2\%, confirms that this state is loosely bound. The masses of the first two
states from the relativistic framework look more reasonable. We mention that 
a recent study of the oblique parameters $S$ and $T$ has permitted heavy resonances to be 
heavier than 3 TeV \cite{Rosell:2023xlf}. Equations~(\ref{ga2}) and 
(\ref{ga3}) hint that the widths of these bound states are of the same order of magnitude,
so Eq.~(\ref{per}) indicates a tiny contribution  from the $n=2$ state at $10^{-3}$ level to 
the Higgs boson production via gluon fusion. We conclude that the $S$ contributions 
are negligible compared with the SM one. It is thus likely that a fourth 
generation quark as heavy as 2.7 TeV bypasses the constraint of the measured $gg\to H$ cross 
section at the scale $s\sim m_H^2$. 

The same observation holds for the constraint 
on the fourth generation quarks from the data of the Higgs decay into photon pairs. 
The reasoning related to the $H \to \gamma \gamma$ decay proceeds in a similar way. One just 
replaces the effective coupling $g_{ggS}$ in Eq.~(\ref{eff}) by $g_{\gamma\gamma S}$, and the 
constant $3/2$ on the right-hand side of Eq.~(\ref{eff}) by $1/2$, which takes into account the 
color factor for the quark loop and the electric charge of a top quark. We then estimate
the suppression factor on the $S$ contribution relative to the top-quark one, 
\begin{eqnarray}
\left(2\frac{m_H\Gamma_S}{m_S^2}\right)^2\approx  10^{-2}.
\end{eqnarray}
That is, the contribution from the $\bar b'b'$ bound state to the $H \to \gamma \gamma$ 
decay is also negligible.

It is impossible to detect a $t'$ quark with a 
mass as high as 200 TeV in the foreseeable future. To detect a $b'$ quark, the gluon fusion 
into a $\bar b'b'$ resonance of mass about 3.2 TeV may not be efficient 
owing to the small gluon distribution functions at large parton momenta. Instead, the fusion 
process $qq\to WW,ZZ\to S$ \cite{CDH} is more promising, whose cross section is enhanced by 
the quark distribution functions. Another promising channel is the $W$-boson mediated single 
$b'$ quark production associated with a top quark and a light quark, such as 
$dg\to u\bar tb'$. It gains the power enhancement with one fewer virtual weak boson by paying 
the price of having a smaller gluon distribution function. Presuming that $b'$ decays into 
$tW$ dominantly, one can search for an excess of $t\bar tW$ final states 
\cite{CMS:2022tkv,ATLAS:2023gon}. The analysis 
is analogous to the search of vector-like heavy quarks \cite{Canbay:2023vmj}, and the currently 
available strategies work. Another simpler single $b'$ quark production from the $ug\to W^+b'$ 
process may be attempted, which, however, suffers the uncertain suppression of the diminishing 
$4\times 4$ CKM matrix element $V_{ub'}$.

\section{CONCLUSION}

After accumulating sufficient clues in our previous studies that the scalar sector 
of the SM can be stringently constrained and there might be only three fundamental 
parameters from the gauge groups, we delved into the sequential fourth generation model
as a natural extension of the SM. It has been demonstrated that the fourth generation
quark masses can be predicted in the dispersive analyses of neutral quark state mixing 
involving a heavy quark. The idea is to treat the dispersion relations obeyed by 
the mixing observables as inverse problems with the initial conditions from the box-diagram 
contributions in the interval between the quark-level and hadronic thresholds.  
A heavy quark must take a specific mass in order to ensure a physical solution 
for the mixing observable to be invariant under the arbitrary scaling of the heavy quark 
mass in the dispersive integrals. We first worked on the mixing mediated by the 
$db$, $sb$ and $bb$ channels, and showed that the roots of the heavy-quark mass $m_Q$ 
corresponding to the first (second, third) peaks of the $bb$ ($sb$, $db$) contributions, 
with the inputs of the typical strange- and bottom-quark masses, coincide around 
$m_Q\approx 173$ GeV. This outcome, highly nontrivial from the three independent channels 
and in agreement with the measured top quark mass, affirms our claim that the scalar 
interaction introduced to couple different generations in the SM is not discretionary.

Encouraged by the successful explanation of the top quark mass, we applied the 
formalism to the predictions for the fourth generation quark masses. The perturbative inputs 
to the dispersion relations come from the same box diagrams involving multiple intermediate 
channels, i.e., the $ut$ and $ct$ channels in the $b'$ case, and the $db'$, $sb'$ and $bb'$ 
ones in the $t'$ case. As expected, we solved for the common masses $m_{b'}=(2.7\pm 0.1)$ TeV 
and $m_{t'}\approx 200$ TeV from the above channels, which should be solid and convincing. 
Such superheavy quarks with the huge Yukawa couplings form bound states. 
The contributions from the $\bar b'b'$ scalars to Higgs boson production 
via gluon fusion were assessed in an effective approach. Employing the eigenfunctions for scalar 
bound states in a Yukawa potential available in the literature, we calculated the widths 
appearing in the Breit-Wigner propagator associated with the scalars. We further fixed
the relevant effective couplings for the gluon-gluon-scalar vertices and for the new scalar 
transition to a Higgs boson. The new scalar contributions at the scale of the Higgs boson mass 
turned out to be of $O(10^{-3})$ of the top-quark one in the SM at most, and is 
negligible. This estimate illustrated why these superheavy quarks could bypass the 
current experimental constraints from Higgs boson production via gluon fusion and decay to 
photon pairs, and why one should continue the search for fourth generation $b'$ quarks 
or their resonances at the (high-luminosity) large hadron collider.

\appendix

\section{Derivation OF THE DISPERSION RELATION AND ITS SOLUTION}

We recapture the derivation of the dispersion relation in Eq.~(\ref{it}) and
of its solution in Eq.~(\ref{d2}) for a self-contained presentation.
Express the mixing amplitude for the neutral states $Q\bar u$ and $\bar Qu$ as
\begin{eqnarray}
{\cal A}(m_Q)&=&\sum_{i,j}\lambda_i\lambda_j\left[\left(M_{ij}(m_Q)+i\Gamma_{ij}(m_Q)\right)
\bar v_{\bar u}\gamma^\mu(1-\gamma_5)u_Q
\bar u_u\gamma_{\mu}(1-\gamma_5) v_{\bar Q}\right.\nonumber\\
& &\left.+\left(M'_{ij}(m_Q)+i\Gamma'_{ij}(m_Q)\right)\bar v_{\bar u}(1-\gamma_5)u_Q
\bar u_u(1-\gamma_5) v_{\bar Q}\right],
\end{eqnarray}
where $m_Q$ is the mass of the heavy quark $Q$, the light quark $u$ is assumed to be 
massless for simplicity, $\lambda_i\equiv V^*_{Qi}V_{ui}$ are the products of the 
CKM matrix elements, $u_Q$, $v_{\bar Q}$, $\cdots$ represent
the quark spinors, the first (second) term on the right-hand 
side denotes the $(V-A)(V-A)$ ($(S-P)(S-P)$) structure, and $M^{(\prime)}_{ij}$ 
($\Gamma^{(\prime)}_{ij}$) collects the real (imaginary) piece of the amplitude.

Since the last line in Eq.~(\ref{asy}) contains an odd power of $1/m_Q$,  
the construction of a dispersion relation should begin with the contour integration in the 
complex $m$ plane, instead of the $m^2$ plane, which possesses different branching cuts. 
The designated contour has been described in Sec.~II, and depicted in Fig.~\ref{fig1}. 
As stated in Sec.~II, we focus on the $(V-A)(V-A)$ contribution, and 
consider the contour integrations of the modified amplitudes 
\begin{eqnarray}
\Pi_{ij}(m)\equiv F_{ij}(m)\left[M_{ij}(m)+i\Gamma_{ij}(m)\right],
\end{eqnarray}
where the functions $F_{ij}(m)$ have been defined in Eq.~(\ref{mi}),
\begin{eqnarray}
F_{db}(m)&=&\frac{m^4}{(m^2-m_b^2)^2},\nonumber\\
F_{sb}(m)&=&\frac{m^4}{[m^2-(m_b+m_s)^2]^2\sqrt{m^2-(m_b-m_s)^2}^3},
\nonumber\\
F_{bb}(m)&=&\frac{m}{m^2-4m_b^2}.\label{miq}
\end{eqnarray}
The analytical properties of the above amplitudes in the $m$ plane have been discussed 
in Sec.~II.

We have the identity
\begin{eqnarray}
\frac{1}{2\pi i}\oint \frac{\Pi_{ij}(m)}{m-m_Q}dm=0,\label{con}
\end{eqnarray}
which vanishes, for the contour encloses unphysical regions and residues of potential poles 
at low $m$ have been suppressed by $F_{ij}(m)$ in Eq.~(\ref{miq}). The 
contribution along the small clockwise circle in Fig.~\ref{fig1} yields 
${\rm Re}\Pi_{ij}(m_Q)$, and those from the four horizontal sections lead to the 
dispersive integrals of ${\rm Im}\Pi_{ij}(m)$. Equation~(\ref{con}) is rewritten as
\begin{eqnarray}
{\rm Re}\Pi_{ij}(m_Q)=\frac{1}{\pi}\int_{M_{ij}}^R \frac{{\rm Im}\Pi_{ij}(m)}{m-m_Q}dm
-\frac{1}{\pi}\int^{-M_{ij}}_{-R} \frac{{\rm Im}\Pi_{ij}(m)}{m-m_Q}dm
+\frac{1}{2\pi i}\int_{C_R} \frac{\Pi_{ij}^{\rm box}(m)}{m-m_Q}dm,\label{ij}
\end{eqnarray}
with the hadronic threshold $M_{ij}$. The unknown function ${\rm Im}\Pi_{ij}(m)$ acquires
nonperturbative contributions from the small $m$ region. The integrand $\Pi_{ij}(m)$, 
taking values along the large counterclockwise circle $C_R$, can be reliably replaced by the 
perturbative one $\Pi_{ij}^{\rm box}(m)$.

The real part ${\rm Re}\Pi_{ij}^{\rm box}$ and the imaginary part ${\rm Im}\Pi_{ij}^{\rm box}$ 
of the box-diagram contribution respect the dispersion relation
\begin{eqnarray}
{\rm Re}\Pi_{ij}^{\rm box}(m_Q)=\frac{1}{\pi}
\int_{m_{ij}}^R \frac{{\rm Im}\Pi_{ij}^{\rm box}(m)}{m-m_Q}dm
-\frac{1}{\pi}\int^{-m_{ij}}_{-R} \frac{{\rm Im}\Pi_{ij}^{\rm box}(m)}{m-m_Q}dm
+\frac{1}{2\pi i}\int_{C_R} \frac{\Pi_{ij}^{\rm box}(m)}{m-m_Q}dm,\label{ope}
\end{eqnarray}
because of the analyticity, $m_{ij}$ in the first two integrals on the right-hand 
side being the quark-level threshold. We equate ${\rm Re}\Pi_{ij}(m_Q)$ and 
${\rm Re}\Pi_{ij}^{\rm box}(m_Q)$, i.e., 
Eqs.~(\ref{ij}) and (\ref{ope}) at large enough $m_Q\gg M_{ij}$, where perturbative
evaluations are reliable, arriving at
\begin{eqnarray}
\int_{M_{ij}}^R\frac{{\rm Im}\Pi_{ij}(m)}{m-m_Q}dm-
\int^{-M_{ij}}_{-R}\frac{{\rm Im}\Pi_{ij}(m)}{m-m_Q}dm=
\int_{m_{ij}}^R\frac{{\rm Im}\Pi_{ij}^{\rm box}(m)}{m-m_Q}dm-
\int^{-m_{ij}}_{-R}\frac{{\rm Im}\Pi_{ij}^{\rm box}(m)}{m-m_Q}dm.\label{ij1}
\end{eqnarray} 
The contributions from the large circle $C_R$ on the two sides have been canceled. 
The modified amplitudes ${\rm Im}\Pi_{ij}(m)$ and ${\rm Im}\Pi_{ij}^{\rm box}(m)$ are 
even functions of $m$. Hence, we apply the variable change $m\to -m$ to the second integrals 
on both sides of Eq.~(\ref{ij1}), which then reduces to Eq.~(\ref{it0}) in the standard form
with the integration variable $m^2$.

%\begin{eqnarray}
%& &(-i)\frac{1}{4}\left(-i\frac{g}{2\sqrt{2}}\right)^4\int \frac{d^4l}{(2\pi)^4}
%\frac{\bar u_q(0)\gamma_\nu(1-\gamma_5)i(\not p_Q-\not l+m_i)\gamma_\mu(1-\gamma_5) u_Q(p_Q)
%\bar v_{\bar q}(0)\gamma_{\mu'}(1-\gamma_5)(-i)(\not l-m_j)
%\gamma_{\nu'}(1-\gamma_5) v_{\bar Q}(p_Q)}
%{(l^2-m_j^2)[(p_Q-l)^2-m_i^2]}\nonumber\\
%& &\times(-i)^2\frac{g^{\nu\nu'}-(p_Q-l)^\nu(p_Q-l)^{\nu'}/m_W^2}{(p_Q-l)^2-m_W^2}
%\frac{g^{\mu\mu'}-l^\mu l^{\mu'}/m_W^2}{l^2-m_W^2}
%&\equiv&,
%\end{eqnarray}

The variable changes $m_Q^2-m_{ij}^2=u\Lambda$ and $m^2-m_{ij}^2=v\Lambda$, $\Lambda$
being an arbitrary scale, turn Eq.~(\ref{it}) into
\begin{eqnarray}
\int_{0}^\infty dv\frac{\Delta\rho_{ij}(v)}{u-v}=0.\label{i2}
\end{eqnarray}
Since $\Delta\rho_{ij}(v)$ diminishes at large $v$; namely, the major contribution to 
Eq.~(\ref{i2}) comes from the region with finite $v$, we expand Eq.~(\ref{i2}) into a 
power series in $1/u$ for sufficiently large but still arbitrary $u$ by inserting 
$1/(u-v)=\sum_{i=k}^\infty v^{k-1}/u^k$. Equation~(\ref{i2}) thus demands a vanishing 
coefficient for every power of $1/u$. We start with the case with $N$ vanishing coefficients,
\begin{eqnarray}
\int_{0}^\infty dvv^{k-1}\Delta\rho_{ij}(v)=0,\;\;\;\;k=1,2,3\cdots,N,\label{i3}
\end{eqnarray}
where $N$ will be extended to infinity eventually. The first $N$ generalized Laguerre  
polynomials $L_{0}^{(\alpha)}(v)$, $L_{1}^{(\alpha)}(v)$, $\cdots$, $L_{N-1}^{(\alpha)}(v)$ 
are composed of the terms $1$, $v$, $\cdots$, $v^{N-1}$ appearing in the above expressions. 
Therefore, Eq.~(\ref{i3}) implies an expansion of $\Delta\rho_{ij}(v)$ in terms of 
$L_k^{(\alpha)}(v)$ with degrees $k$ not lower than $N$,
\begin{eqnarray}
\Delta \rho_{ij}(v)=\sum_{k=N}^{N_{ij}} a^{(k)}_{ij}v^{\alpha_{ij}} e^{-v}
L_{k}^{(\alpha_{ij})}(v),\;\;\;\;N_{ij}>N,\label{d0}
\end{eqnarray}
owing to the orthogonality of the polynomials, in which $a^{(k)}_{ij}$ represent a set of 
unknown coefficients. The index $\alpha_{ij}$ describes the behavior of $\Delta \rho_{ij}(v)$ 
around $v\sim 0$. The highest degree $N_{ij}$ can be fixed in principle by the initial 
condition $-{\rm Im}\Pi_{ij}^{\rm box}(v)$ of $\Delta \rho_{ij}(v)$ in the interval 
$(0,(M_{ij}^2-m_{ij}^2)/\Lambda)$ of $v$. Because $-{\rm Im}\Pi_{ij}^{\rm box}(v)$ is a 
smooth function, $N_{ij}$ needs not be infinite.

%\begin{eqnarray}
%L_j^{(\alpha)}(v)\approx j^{\alpha/2}v^{-\alpha/2}e^{v/2}J_\alpha(2\sqrt{jv}),\label{Ln}
%\end{eqnarray}

A generalized Laguerre polynomial takes the approximate form for large $k$,
$L_k^{(\alpha)}(v)\approx k^{\alpha/2}v^{-\alpha/2}e^{v/2}J_\alpha(2\sqrt{kv})$ \cite{BBC},
up to corrections of $1/\sqrt{k}$. Equation~(\ref{d0}) becomes
\begin{eqnarray}
\Delta \rho_{ij}(m)
\approx\sum_{k=N}^{N_{ij}} a^{(k)}_{ij}\sqrt{\frac{k(m^2-m_{ij}^2)}{\Lambda}}^{\alpha_{ij}} 
e^{-(m^2-m_{ij}^2)/(2\Lambda)}
J_{\alpha_{ij}}\left(2\sqrt{\frac{k(m^2-m_{ij}^2)}{\Lambda}}\right),
\label{d1}
\end{eqnarray}
where $v=(m^2-m_{ij}^2)/\Lambda$ has been inserted. Defining the 
scaling variable $\omega\equiv\sqrt{N/\Lambda}$, we have the approximation 
$N_{ij}/\Lambda=\omega^2[1+(N_{ij}-N)/N]\approx \omega^2$ for finite $N_{ij}-N$. 
The common Bessel functions $J_{\alpha_{ij}}(2\sqrt{k(m^2-m_{ij}^2)/\Lambda})\approx 
J_{\alpha_{ij}}(2\omega \sqrt{m^2-m_{ij}^2})$ for $k=N,N+1,\cdots,N_{ij}$ can be factored 
out, such that the unknown coefficients are summed into a single parameter 
$y_{ij}=\sum_{k=N}^{N_{ij}} a^{(k)}_{ij}$. We are allowed to treat $\omega$ as a finite 
variable, though both $N$ and $\Lambda$ can be arbitrarily large. The arbitrariness 
of $\Lambda$, which traces back to that of the large circle radius $R$, goes into the
variable $\omega$. The exponential suppression 
factor $e^{-(m^2-m_{ij}^2)/(2\Lambda)}=e^{-\omega^2 (m^2-m_{ij}^2)/(2N)}$ is further replaced 
by unity for finite $\omega$ and large $N$. Equation~(\ref{d1}) then gives 
the solution in Eq.~(\ref{d2}).

%\begin{eqnarray}
%\Delta \rho_{ij}(m)\approx
%y_{ij}\left(\omega \sqrt{m^2-m_{ij}^2}\right)^{\alpha_{ij}} 
%J_{\alpha_{ij}}\left(2\omega \sqrt{m^2-m_{ij}^2}\right),
%\label{i4}
%\end{eqnarray}

\section*{Acknowledgement}

We thank K.F. Chen, Y.T. Chien, X.G. He, W.S. Hou and P.Q. Hung for stimulating discussions.
This work was supported in part by National Science and Technology Council of the Republic
of China under Grant No. MOST-110-2112-M-001-026-MY3.

%%%%%%%%%%%%%%%%%%%%%%%%%%%%%%%%%%%%%%%%%%%%%%%%%%%%%%%%%%%%%%%%%%%%%%%%%%%%%%%%%%%%%%%%%%%%%%%%%%%%%%%%%%%%%%%%%%%%%%%%%%%%


\begin{thebibliography}{99}


\bibitem{Li:2023dqi}
H.~n.~Li,
%``Dispersive constraints on fermion masses,''
Phys. Rev. D \textbf{107}, no.9, 094007 (2023).
%doi:10.1103/PhysRevD.107.094007
%[arXiv:2302.01761 [hep-ph]].


\bibitem{Li:2023yay}
H.~n.~Li,
%``Dispersive determination of electroweak-scale masses,''
Phys. Rev. D \textbf{108}, no.5, 054020 (2023).
%doi:10.1103/PhysRevD.108.054020
%[arXiv:2304.05921 [hep-ph]].

\bibitem{Li:2023ncg}
H.~n.~Li,
%``Dispersive determination of neutrino mass ordering,''
[arXiv:2306.03463 [hep-ph]].

\bibitem{Li:2020xrz} 
  H.~n.~Li, H.~Umeeda, F.~Xu and F.~S.~Yu,
%``$D$ meson mixing as an inverse problem,''
Phys. Lett. B \textbf{810}, 135802 (2020).

\bibitem{Li:2020fiz}
H.~n.~Li and H.~Umeeda,
%``Vacuum polarization contribution to muon $g-2$ as an inverse problem,''
Phys. Rev. D \textbf{102}, no.9, 094003 (2020).
%doi:10.1103/PhysRevD.102.094003
%[arXiv:2004.06451 [hep-ph]].

\bibitem{Li:2020ejs}
H.~n.~Li and H.~Umeeda,
%``QCD sum rules with spectral densities solved in inverse problems,''
Phys. Rev. D \textbf{102}, 114014 (2020).
%doi:10.1103/PhysRevD.102.114014
%[arXiv:2006.16593 [hep-ph]].

\bibitem{Xiong:2022uwj}
A.~S.~Xiong, T.~Wei and F.~S.~Yu,
%``Inverse Problem Approach for Non-Perturbative QCD: Foundation,''
arXiv:2211.13753 [hep-th].

\bibitem{Belfatto:2023qca}
B.~Belfatto and Z.~Berezhiani,
%``Minimally modified Fritzsch texture for quark masses and CKM mixing,''
[arXiv:2305.00069 [hep-ph]].

\bibitem{Santamaria:1993ah}
A.~Santamaria,
%``Masses, mixings, Yukawa couplings and their symmetries,''
Phys. Lett. B \textbf{305}, 90-97 (1993).
%doi:10.1016/0370-2693(93)91110-9
%[arXiv:hep-ph/9302301 [hep-ph]].

\bibitem{HBH} B.~Holdom, Phys. Rev. Lett. {\bf 57}, 2496 (1986), [Erratum-ibid. 58, 177 (1987)];
W.~A.~Bardeen, C.~T.~Hill and M.~Lindner, Phys. Rev. D {\bf 41}, 1647 (1990); C.~T.~Hill,
M.~A.~Luty and E.~A.~Paschos, Phys. Rev. D {\bf 43}, 3011 (1991); T.~Elliott and S.~F.~King,
Phys. Lett. B {\bf 283}, 371 (1992).


\bibitem{Mimura:2012vw}
Y.~Mimura, W.~S.~Hou and H.~Kohyama,
%``Bootstrap dynamical symmetry breaking with strong Yukawa coupling,''
JHEP \textbf{11}, 048 (2013).
%doi:10.1007/JHEP11(2013)048
%[arXiv:1206.6063 [hep-ph]].

\bibitem{HOS} S.~W.~Ham, S.~K.~Oh and D.~Son, Phys. Rev. D {\bf 71}, 015001 (2005);
%[arXiv:hep-ph/0411012]; 
M.~S.~Carena, A.~Megevand, M.~Quiros and C.~E.~M. Wagner, Nucl. Phys. B {\bf 716}, 319 (2005); 
%[arXiv:hep-ph/0410352]; 
R.~Fok and G.~D.~Kribs, Phys. Rev. D {\bf 78}, 075023 (2008) 
%[arXiv:0803.4207 [hep-ph]]; 
%\cite{Kikukawa:2009mu}
%\bibitem{Kikukawa:2009mu}
Y.~Kikukawa, M.~Kohda and J.~Yasuda,
%``The Strongly coupled fourth family and a first-order electroweak phase transition. I. Quark sector,''
Prog. Theor. Phys. \textbf{122}, 401-426 (2009).
%doi:10.1143/PTP.122.401
%[arXiv:0901.1962 [hep-ph]].

%\bibitem{Hung:1997zj}
%P.~Q.~Hung,
%%``Minimal SU(5) resuscitated by longlived quarks and leptons,''
%Phys. Rev. Lett. \textbf{80}, 3000-3003 (1998).
%doi:10.1103/PhysRevLett.80.3000
%[arXiv:hep-ph/9712338 [hep-ph]].


\bibitem{Hou:2008xd}
W.~S.~Hou,
%``Source of CP Violation for the Baryon Asymmetry of the Universe,''
Chin. J. Phys. \textbf{47}, 134 (2009).
%[arXiv:0803.1234 [hep-ph]].
%187 citations counted in INSPIRE as of 22 Jun 2023


\bibitem{Chen:2012wz}
N.~Chen and H.~J.~He,
%``LHC Signatures of Two-Higgs-Doublets with Fourth Family,''
JHEP \textbf{04}, 062 (2012);
%doi:10.1007/JHEP04(2012)062
%[arXiv:1202.3072 [hep-ph]].
%\cite{Eberhardt:2012gv}
%\bibitem{Eberhardt:2012gv}
O.~Eberhardt, G.~Herbert, H.~Lacker, A.~Lenz, A.~Menzel, U.~Nierste and M.~Wiebusch,
%``Impact of a Higgs boson at a mass of 126 GeV on the standard model with three and four fermion generations,''
Phys. Rev. Lett. \textbf{109}, 241802 (2012);
%doi:10.1103/PhysRevLett.109.241802
%[arXiv:1209.1101 [hep-ph]].
%\cite{Djouadi:2012ae}
%\bibitem{Djouadi:2012ae}
A.~Djouadi and A.~Lenz,
%``Sealing the fate of a fourth generation of fermions,''
Phys. Lett. B \textbf{715}, 310-314 (2012);
%doi:10.1016/j.physletb.2012.07.060
%[arXiv:1204.1252 [hep-ph]].
%\cite{Kuflik:2012ai}
%\bibitem{Kuflik:2012ai}
E.~Kuflik, Y.~Nir and T.~Volansky,
%``Implications of Higgs searches on the four generation standard model,''
Phys. Rev. Lett. \textbf{110}, no.9, 091801 (2013).
%doi:10.1103/PhysRevLett.110.091801
%[arXiv:1204.1975 [hep-ph]].


\bibitem{He:2001tp}
H.~J.~He, N.~Polonsky and S.~f.~Su,
%``Extra families, Higgs spectrum and oblique corrections,''
Phys. Rev. D \textbf{64}, 053004 (2001).
%doi:10.1103/PhysRevD.64.053004
%[arXiv:hep-ph/0102144 [hep-ph]].

\bibitem{Hung:2009hy}
P.~Q.~Hung and C.~Xiong,
%``Renormalization Group Fixed Point with a Fourth Generation: Higgs-induced Bound States and Condensates,''
Nucl. Phys. B \textbf{847}, 160-178 (2011).
%doi:10.1016/j.nuclphysb.2011.01.025
%[arXiv:0911.3890 [hep-ph]].


\bibitem{Enkhbat:2011vp}
T.~Enkhbat, W.~S.~Hou and H.~Yokoya,
%``Early LHC Phenomenology of Yukawa-bound Heavy $Q\bar{Q}$ Mesons,''
Phys. Rev. D \textbf{84}, 094013 (2011).
%doi:10.1103/PhysRevD.84.094013
%[arXiv:1109.3382 [hep-ph]].


\bibitem{Hung:2010xh}
P.~Q.~Hung and C.~Xiong,
%``Dynamical Electroweak Symmetry Breaking with a Heavy Fourth Generation,''
Nucl. Phys. B \textbf{848}, 288-302 (2011).
%doi:10.1016/j.nuclphysb.2011.02.018
%[arXiv:1012.4479 [hep-ph]].

\bibitem{Das:2017mnu}
D.~Das, A.~Kundu and I.~Saha,
%``Higgs data does not rule out a sequential fourth generation with an extended scalar sector,''
Phys. Rev. D \textbf{97}, no.1, 011701 (2018).
%doi:10.1103/PhysRevD.97.011701
%[arXiv:1707.03000 [hep-ph]].

\bibitem{Li:2022jxc}
H.~n.~Li,
%``Dispersive analysis of neutral meson mixing,''
Phys. Rev. D \textbf{107}, no.5, 054023 (2023).
%doi:10.1103/PhysRevD.107.054023
%[arXiv:2208.14798 [hep-ph]].

\bibitem{Petrov:1997ch}
A.~A.~Petrov,
%``On dipenguin contribution to D0 - anti-D0 mixing,''
Phys. Rev. D \textbf{56}, 1685 (1997).
%Phys. Rev. D \textbf{56}, 1685-1687 (1997).
%doi:10.1103/PhysRevD.56.1685
%[arXiv:hep-ph/9703335 [hep-ph]].

\bibitem{Cheng} 
H.~Y.~Cheng, Phys. Rev. D {\bf 26}, 143 (1982).

\bibitem{BSS}
A.~J.~Buras, W.~Slominski and H.~Steger,
Nucl. Phys. {\bf B245}, 369 (1984).
%369-398



\bibitem{Buchalla:1995vs}
G.~Buchalla, A.~J.~Buras and M.~E.~Lautenbacher,
%``Weak decays beyond leading logarithms,''
Rev. Mod. Phys. \textbf{68}, 1125-1144 (1996).
%doi:10.1103/RevModPhys.68.1125
%[arXiv:hep-ph/9512380 [hep-ph]].

\bibitem{SVZ} M.~A.~Shifman, A.~I.~Vainshtein and V.~I.~Zakharov, Nucl. Phys. B {\bf 147}, 385 (1979);
B {\bf 147}, 448 (1979).

\bibitem{Zhong:2021epq}
T.~Zhong, Z.~H.~Zhu, H.~B.~Fu, X.~G.~Wu and T.~Huang,
%``Improved light-cone harmonic oscillator model for the pionic leading-twist distribution amplitude,''
Phys. Rev. D \textbf{104}, 016021 (2021).
%doi:10.1103/PhysRevD.104.016021
%[arXiv:2102.03989 [hep-ph]].

\bibitem{PDG}
R.L. Workman et al. (Particle Data Group), Prog. Theor. Exp. Phys. 2022, 083C01 (2022).
%  M. Tanabashi {\it et al}. (Particle Data Group), Phys. Rev. D {\bf 98}, 030001 (2018).


\bibitem{FermilabLattice:2018est}
A.~Bazavov \textit{et al.} [Fermilab Lattice, MILC and TUMQCD],
%``Up-, down-, strange-, charm-, and bottom-quark masses from four-flavor lattice QCD,''
Phys. Rev. D \textbf{98}, no.5, 054517 (2018).
%doi:10.1103/PhysRevD.98.054517
%[arXiv:1802.04248 [hep-lat]].

\bibitem{Deur:2016tte}
A.~Deur, S.~J.~Brodsky and G.~F.~de Teramond,
%``The QCD Running Coupling,''Prog. Part. Nuc. Phys. 90 1 (2016)
Prog. Part. Nucl. Phys. \textbf{90}, 1 (2016).
%doi:10.1016/j.ppnp.2016.04.003
%[arXiv:1604.08082 [hep-ph]].


\bibitem{Hou:2012df}
W.~S.~Hou,
%``Searching for new heavy chiral quark pairs via their annihilation to multiple vector bosons,''
Phys. Rev. D \textbf{86}, 037701 (2012).
%doi:10.1103/PhysRevD.86.037701
%[arXiv:1206.1453 [hep-ph]].

\bibitem{Napsuciale:2021qtw}
M.~Napsuciale and S.~Rodr\'\i{}guez,
%``Bound states of the Yukawa potential from hidden supersymmetry,''
PTEP \textbf{2021}, no.7, 073B03 (2021).
%doi:10.1093/ptep/ptab070
%[arXiv:2102.07160 [hep-ph]].

\bibitem{TH}
L.~H.~Thomas, Phys. Rev. {\bf 47}, 903 (1935).

\bibitem{Ikhdair:2012zz}
S.~M.~Ikhdair,
%``Approximate kappa-state solutions to the Dirac-Yukawa problem based on the spin and pseudospin symmetry,''
Central Eur. J. Phys. \textbf{10}, 361-381 (2012).
%doi:10.2478/s11534-011-0121-5
%[arXiv:1203.2023 [quant-ph]].

\bibitem{Lansberg:2009xh}
J.~P.~Lansberg and T.~N.~Pham,
%``Effective Lagrangian for Two-photon and Two-gluon Decays of P-wave Heavy Quarkonium chi(c0,2) and chi(b0,2) states,''
Phys. Rev. D \textbf{79}, 094016 (2009).
%doi:10.1103/PhysRevD.79.094016
%[arXiv:0903.1562 [hep-ph]].

\bibitem{GGM}
H.~M.~Georgi, S.~L.~Glashow, M.~E.~Machacek, and D.~V.~Nanopoulos, Phys. Rev.
Lett. {\bf 40}, 692 (1978).

\bibitem{Spira:1995rr}
M.~Spira, A.~Djouadi, D.~Graudenz and P.~M.~Zerwas,
%``Higgs boson production at the LHC,''
Nucl. Phys. B \textbf{453}, 17-82 (1995).
%doi:10.1016/0550-3213(95)00379-7
%[arXiv:hep-ph/9504378 [hep-ph]].

\bibitem{Rosell:2023xlf}
I.~Rosell, A.~Pich and J.~J.~Sanz-Cillero,
%``Heavy resonances and the oblique parameters S and T,''
[arXiv:2309.09741 [hep-ph]].

\bibitem{CDH} R.~N.~Cahn and S.~Dawson, Phys. Lett. {\bf 136B}, 196 (1984)
K.~Hikasa, Phys. Lett. {\bf 164B}, 385 (1985);
G.~Altarelli, B.~Mele and F.~Pitolli, Nucl. Phys. {\bf B287}, 205 (1987);
T.~Han, G.~Valencia and S.~Willenbrock, Phys. Rev. Lett. {\bf 69}, 3274 (1992).


\bibitem{CMS:2022tkv}
A.~Tumasyan \textit{et al.} [CMS],
%``Measurement of the cross section of top quark-antiquark pair production in association with a W boson in proton-proton collisions at $ \sqrt{s} $ = 13 TeV,''
JHEP \textbf{07}, 219 (2023).
%doi:10.1007/JHEP07(2023)219
%[arXiv:2208.06485 [hep-ex]].


\bibitem{ATLAS:2023gon}
 [ATLAS],
%``Measurement of the total and differential cross-sections of $t\bar{t}W$ production in $pp$ collisions at 13 TeV with the ATLAS detector,''
ATLAS-CONF-2023-019.

\bibitem{Canbay:2023vmj}
A.~C.~Canbay and O.~Cakir,
%``Investigating the Single Production of Vector-Like Quarks Decaying into Top Quark and W Boson through Hadronic Channels at the HL-LHC,''
[arXiv:2307.12883 [hep-ph]].

\bibitem{BBC} D.~Borwein, J.~M.~Borwein, R.~E.~Crandall, 
%"Effective Laguerre asymptotics", 
SIAM J. Numer. Anal. {\bf 46}, 3285–3312 (2008).
% doi:10.1137/07068031X.


\end{thebibliography}
\end{document}